\documentclass{emulateapj}
\usepackage{apjfonts}

\newcommand{\rl}{$R_{\rm BLR}$--$L$}
\newcommand{\mlbulge}{$M_{\rm BH} - L_{\rm bulge}$}
\newcommand{\msigma}{$M_{\rm BH}-\sigma_{\star}$}
\newcommand{\mbh}{$M_{\rm BH}$}

\shorttitle{LAMP: H$\beta$ BLR RADII AND BLACK HOLE MASSES}
\shortauthors{BENTZ ET AL.}

\begin{document}

\title{The Lick AGN Monitoring Project: Broad-Line Region Radii and 
Black Hole Masses from Reverberation Mapping of H$\beta$}

\author{ 
Misty~C.~Bentz\altaffilmark{1,2},
Jonelle~L.~Walsh\altaffilmark{1},
Aaron~J.~Barth\altaffilmark{1},
Nairn~Baliber\altaffilmark{3,4},
Nicola~Bennert\altaffilmark{3,5},
Gabriela~Canalizo\altaffilmark{5,6},
Alexei~V.~Filippenko\altaffilmark{7},
Mohan~Ganeshalingam\altaffilmark{7},
Elinor~L.~Gates\altaffilmark{8},
Jenny~E.~Greene\altaffilmark{2,9},
Marton~G.~Hidas\altaffilmark{3,4,10},
Kyle~D.~Hiner\altaffilmark{5,6},
Nicholas~Lee\altaffilmark{7},
Weidong~Li\altaffilmark{7},
Matthew~A.~Malkan\altaffilmark{11},
Takeo~Minezaki\altaffilmark{12},
Yu~Sakata\altaffilmark{12,13},
Frank~J.~D.~Serduke\altaffilmark{7},
Jeffrey~M.~Silverman\altaffilmark{7},
Thea~N.~Steele\altaffilmark{7},
Daniel~Stern\altaffilmark{14},
Rachel~A.~Street\altaffilmark{3,4},
Carol~E.~Thornton\altaffilmark{1},
Tommaso~Treu\altaffilmark{3,15},
Xiaofeng~Wang\altaffilmark{7,16},
Jong-Hak~Woo\altaffilmark{2,3,11}, and
Yuzuru~Yoshii\altaffilmark{12,17}
}

\altaffiltext{1}{Department of Physics and Astronomy,
                 4129 Frederick Reines Hall,
                 University of California,
                 Irvine, CA 92697;
                 mbentz@uci.edu .} 

\altaffiltext{2}{Hubble Fellow.}

\altaffiltext{3}{Physics Department, 
                 University of California, 
                 Santa Barbara, CA 93106.}

\altaffiltext{4}{Las Cumbres Observatory Global Telescope, 
                  6740 Cortona Dr. Ste. 102, 
                  Goleta, CA 93117.}

\altaffiltext{5}{Institute of Geophysics and Planetary Physics,
                 University of California,
                 Riverside, CA 92521.}

\altaffiltext{6}{Department of Physics and Astronomy,
                 University of California,
                 Riverside, CA 92521.}

\altaffiltext{7}{Department of Astronomy, 
                 University of California,
                 Berkeley, CA 94720-3411.}

\altaffiltext{8}{Lick Observatory,
                 P.O. Box 85, 
                 Mount Hamilton, CA 95140.}

\altaffiltext{9}{Princeton University Observatory,
                 Princeton, NJ 08544.}

\altaffiltext{10}{School of Physics A28, 
                  University of Sydney, 
                  NSW 2006, Australia.}

\altaffiltext{11}{Department of Physics and Astronomy, 
                 University of California, 
                 Los Angeles, CA 90024.}

\altaffiltext{12}{Institute of Astronomy, 
                 School of Science, University of Tokyo, 
                 2-21-1 Osawa, Mitaka, Tokyo 181-0015, Japan.}

\altaffiltext{13}{Department of Astronomy,
                  School of Science, University of Tokyo, 
                  7-3-1 Hongo, Bunkyo-ku, Tokyo 113-0033, Japan.}

\altaffiltext{14}{Jet Propulsion Laboratory, 
                  California Institute of Technology, 
                  MS 169-527, 4800 Oak Grove Drive, 
                  Pasadena, CA 91109.}

\altaffiltext{15}{Sloan Fellow, Packard Fellow.}

\altaffiltext{16}{Physics Department and Tsinghua 
                  Center for Astrophysics (THCA), Tsinghua
                  University, Beijing, 100084, China.}

\altaffiltext{17}{Research Center for the Early Universe, 
                  School of Science,
                  University of Tokyo, 7-3-1 Hongo, Bunkyo-ku, 
                  Tokyo 113-0033, Japan.}

\begin{abstract}

We have recently completed a 64-night spectroscopic monitoring
campaign at the Lick Observatory 3-m Shane telescope with the aim of
measuring the masses of the black holes in 12 nearby ($z < 0.05$)
Seyfert~1 galaxies with expected masses in the range $\sim
10^6$--$10^7$~M$_{\odot}$ and also the well-studied nearby active
galactic nucleus (AGN) NGC\,5548.  Nine of the objects in the sample
(including NGC\,5548) showed optical variability of sufficient
strength during the monitoring campaign to allow for a time lag to be
measured between the continuum fluctuations and the response to these
fluctuations in the broad H$\beta$ emission.  We present here the
light curves for all the objects in this sample and the subsequent
H$\beta$ time lags for the nine objects where these measurements were
possible.  The H$\beta$ lag time is directly related to the size of
the broad-line region in AGNs, and by combining the H$\beta$ lag time
with the measured width of the H$\beta$ emission line in the variable
part of the spectrum, we determine the virial mass of the central
supermassive black hole in these nine AGNs.  The absolute calibration
of the black hole masses is based on the normalization derived by
\citeauthor{onken04}, which brings the masses determined by
reverberation mapping into agreement with the local \msigma\
relationship for quiescent galaxies.  We also examine the time lag
response as a function of velocity across the H$\beta$ line profile
for six of the AGNs.  The analysis of four leads to rather ambiguous
results with relatively flat time lags as a function of velocity.
However, SBS\,1116+583A exhibits a symmetric time lag response around
the line center reminiscent of simple models for circularly orbiting
broad-line region (BLR) clouds, and Arp\,151 shows an asymmetric
profile that is most easily explained by a simple gravitational infall
model.  Further investigation will be necessary to fully understand
the constraints placed on physical models of the BLR by the
velocity-resolved response in these objects.

\end{abstract}

\keywords{galaxies: active -- galaxies: nuclei -- galaxies: Seyfert}

\section{Introduction}

Active galactic nuclei (AGNs) have long been known to vary in
luminosity on timescales of years to months or even days
(\citealt{matthews63,smith63}).  Variability has played a central role
in AGN studies.  Combining the physical size constraints set by rapid
variability with the high luminosities of AGNs led to the original
argument that AGNs are powered by accretion onto supermassive black
holes (\citealt{zeldovich64,salpeter64}).  Variability is used as a
reliable method for detecting AGNs in surveys (e.g.,
\citealt{vandenbergh73,heckman76,veron95}), and it is the fundamental
basis upon which rests the technique of measuring black hole masses
known as reverberation mapping (\citealt{blandford82,peterson93}).

Reverberation mapping is the most successful method employed for
measuring the mass of the central black hole in broad emission line
AGNs.  Rather than relying on spatially resolved observations, as do
most studies of black holes in nearby quiescent galaxies,
reverberation mapping resolves the influence of the black hole in the
time domain through spectroscopic monitoring of the continuum flux
variability and the delayed response, or ``echo,'' in the broad
emission line flux.  The time lag between these changes, $\tau$,
depends on the light-travel time across the broad-line region (BLR)
and is on the order of light days for nearby Seyfert galaxies,
corresponding to spatial scales of $\sim 0.001$\,pc.  Combining the
radius of the BLR, $c \tau$, with the velocity width, $v$, of the
broad emission line gives the virial mass of the central black hole
via the simple gravitational relation $M = c \tau v^2 /G$ (neglecting
a factor of order unity).

To date, successful reverberation-mapping studies have been carried
out for approximately 36 active galaxies (compiled by
\citealt{peterson04,peterson05} with additions by
\citealt{bentz06b,denney06,bentz07,grier08}, and \citealt{denney09}).
One of the most important results to come from reverberation mapping
is the detection of a correlation between the BLR radius and the
luminosity of the AGN, the \rl\ relationship
(\citealt{koratkar91b,kaspi00,kaspi05,bentz06a,bentz09b}).  Combining
the \rl\ relationship with the simple virial mass equation results in
an extremely powerful tool for estimating black hole masses in
broad-lined AGNs from a single epoch of spectroscopy and two simple
spectral measurements: the velocity width of a broad emission line,
and the continuum luminosity as a proxy for the radius of the BLR.
The \rl\ relationship is therefore fundamental to all secondary
techniques used to estimate black hole masses in AGNs (e.g.,
\citealt{laor98,wandel99,mclure02,vestergaard06}), and as such,
current studies of black holes in AGNs rest upon the calibration
provided by the reverberation mapping sample (e.g.,
\citealt{onken04,collin06,mcgill08}).

The vast majority of reverberation experiments have investigated black
holes with masses in the range $10^7$--$10^9$~M$_{\odot}$.  Studies of
lower-mass black holes have largely been restricted by the lower
luminosities associated with smaller AGNs, and the few studies that
have been carried out have large measurement uncertainties.  It is
particularly important to have the correct calibration for AGNs in the
mass range of $10^6$--$10^7$~M$_{\odot}$, as they are at the peak of
the local black hole mass distribution function (e.g.,
\citealt{greene07}).  In particular, AGNs in this mass range may
provide strong constraints on the mass accretion history of the
Universe through the coupling of the central black hole and the host
galaxy, as evidenced by the relationship between black hole mass and
bulge luminosity (the \mlbulge\ relationship; e.g.,
\citealt{magorrian98,marconi03,bentz09a}) and the relationship between
black hole mass and bulge stellar velocity dispersion (the \msigma\
relationship; \citealt{ferrarese00,gebhardt00,tremaine02}).

With the goal of extending the range of masses probed by reverberation
studies, we have carried out a 64-night spectroscopic monitoring
campaign with the Lick Observatory 3-m Shane telescope, targeting AGNs
having expected black hole masses in the range $\sim
10^6$--$10^7$~M$_{\odot}$.  We report here the H$\beta$ light curves
and reverberation analysis for the entire sample of 13 AGNs included
in the Lick AGN Monitoring Project (LAMP).  For those objects with
significant correlations between the H$\beta$ and continuum light
curves, of which there were nine, we quantify the time lag between the
variations in the light curves and present the derived black hole
masses.  We also investigate the time lag behavior as a function of
velocity across the H$\beta$ line profile for six of the AGNs.  We
have previously published the H$\beta$ results for one of the objects,
Arp\,151 (Mrk\,40; \citealt{bentz08}, hereafter Paper I), and here we
give an update to the results for Arp\,151 based on slight
modifications to the data processing, to be consistent with all the
results presented here.  The small changes to the measured time lag
and derived black hole mass for Arp\,151 are not significant.

\section{Observations}

Details of the target selection and the photometric monitoring
campaign are presented by \citeauthor{walsh09} (2009, hereafter Paper
II).  In short, the sample of AGNs chosen for this study is listed in
Table~\ref{table:objects} and is comprised of 12 nearby ($z<0.05$)
AGNs with estimated black hole masses (based on single-epoch
spectroscopy) in the range $\sim 10^6$--$10^7$~M$_{\odot}$, expected
H$\beta$ lags between 5--20 days, and relatively strong broad-line
components to their H$\beta$ lines. Also included as a ``control
object'' is NGC\,5548, which has 14 years of previous
reverberation-mapping data and a well-determined black hole mass of
$6.54^{+0.26}_{-0.25} \times 10^7$~M$_{\odot}$ (\citealt{bentz07} and
references therein).  Inclusion of NGC\,5548 adds extra value to our
sample by allowing a direct comparison of our results with those of
previous reverberation mapping experiments.

\subsection{Photometry}

Broad-band Johnson $B$ and $V$ monitoring of all 13 AGNs in the sample
was carried out at four telescopes: the 30-inch robotic Katzman
Automatic Imaging Telescope (KAIT), the 2-meter Multicolor Active Galactic
Nuclei Monitoring telescope, the Palomar 60-inch telescope, and the
32-inch Tenagra II telescope.  The details of the photometric
monitoring are described in Paper II, but we include a summary here.

Each of the four telescopes was responsible for monitoring a subset of
the sample.  Twice-weekly observations of the targets began in early
February 2008, but was increased to nightly monitoring beginning the
evening of 2008 March 17 (UT, both here and throughout), about one
week before the spectroscopic monitoring began on 2008 March 25.  The
photometric light curves mainly follow variations in the continuum
flux, and the response of the broad emission lines is delayed relative
to changes in the continuum.  By starting the photometric monitoring
early, we hoped to ensure that all events at the beginning of the
spectroscopic light curves would have associated events in the
photometric light curves.

The images were reduced following standard techniques. The fluxes of
the AGNs were measured through circular apertures as described in
Paper II and differential photometry was obtained relative to stars
within the fields, which themselves were calibrated to
\citet{landolt92} standard stars.  A simple model of the host galaxy
surface brightness was subtracted from each of the AGN images to help
compensate for the diluting contribution from host-galaxy starlight.
The models did not include a bulge component due to the lack of
spatial resolution in the ground-based images, and so represent a
lower limit to the true host-galaxy contribution.  As we are
interested here in relative flux changes, the absolute scaling of the
AGN flux in the photometry is not important for the results described
in this work.

Flux uncertainties were determined through two methods and the larger
uncertainty contribution was adopted for each datum.  In general, the
flux errors from photon statistics were not large enough to account
for the overall behavior of the light curves, and instead the
uncertainty determined from the average difference between closely
spaced pairs of points in each light curve was adopted.  The
exceptions were generally nights with poor weather conditions, where
the photon-counting statistics provided a larger flux uncertainty.
The $B$- and $V$-band light curves for each of the 13 AGNs are
tabulated in Paper II.

\subsection{Spectroscopy}

Our spectroscopic campaign was carried out over 64 mostly contiguous
nights at the Lick Observatory 3-m Shane telescope between 2008 March
25 and June 1.  We used the Kast dual spectrograph but restricted our
observations to the red-side CCD\footnote{Shortly before our
spectroscopic campaign began, the blue-side CCD in the Kast
spectrograph failed and was replaced by a temporary CCD with a much
lower quantum efficiency.  Rather than extend our exposure times and
decrease the sample of target AGNs, we opted to use only the red-side
CCD.}  and employed the 600~lines~mm$^{-1}$ grating with spectral
coverage over the range 4300--7100\,\AA, giving a nominal resolution
of 2.35\,\AA\,pix$^{-1}$ in the dispersion direction and
0\farcs78\,pix$^{-1}$ in the spatial direction.  Spectra were obtained
through a 4\arcsec-wide slit at fixed position angles for each of the
objects (as listed in Table~\ref{table:observ}).  A fixed position
angle for each individual object is important to mitigate any apparent
variability due to different contributions of starlight from
structures within the host galaxy.  The position angles were set to
match the average parallactic angle expected for each of the objects
throughout the length of the spectroscopic campaign, in an attempt to
lessen the effect of atmospheric dispersion \citep{filippenko82}.
The number of nights on which spectra were obtained for each of the
objects in our sample ranged from 43 to 51, with an average of 47,
which is fairly typical given the historic data on spring observing
conditions at Lick Observatory\footnote{See
http://mthamilton.ucolick.org/techdocs/MH\_weather/obstats/ for
average historic weather records for Lick Observatory.}.

Exposure times, average airmass, and typical signal-to-noise ratio per
pixel in the continuum are also listed in Table~\ref{table:observ}.
The two-dimensional spectroscopic images were reduced with
IRAF\footnote{IRAF is distributed by the National Optical Astronomy
Observatories, which are operated by the Association of Universities
for Research in Astronomy, Inc., under cooperative agreement with the
National Science Foundation.} and an extraction width of 13 pixels (9
pixels for MCG-06-30-15, to avoid a nearby star) was applied,
resulting in spectra with a 10\farcs1 (7\farcs0) extraction.  Sky
regions were included on either side of the extracted regions, of
width 6 pixels and beginning at a distance of 19 pixels to avoid the
vast majority of contribution from the extended host galaxies.  Flux
calibrations were determined from nightly spectra of standard stars,
which typically included Feige\,34 and BD+284211.

To mitigate the effects of slit losses and variable seeing and
transparency, a final, internal calibration of the spectra is
required.  We employed the spectral scaling algorithm of
\citet{vangroningen92} to scale the total flux of the narrow
[\ion{O}{3}] $\lambda \lambda$ 4959,~5007 doublet in each spectrum to
match the [\ion{O}{3}] flux in a reference spectrum created from the
mean of the spectra obtained for each object. This method accounts for
differences in the overall flux scale, as well as small wavelength
shifts and small differences in spectral resolution due to variable
seeing, and has been shown to result in spectrophotometric accuracies
of $\sim 2$\% \citep{peterson98a}.  The adopted absolute scaling of
the [\ion{O}{3}] $\lambda 5007$ line for each object is listed in
Table~\ref{table:o3flux}, along with spectrophotometric [\ion{O}{3}]
$\lambda 5007$ fluxes from the literature for comparison.  From the
available information, it was determined that the night of 2008 April
10 was the only steadily photometric night of the campaign, and
provides the absolute [\ion{O}{3}] scaling for all the objects in our
sample, with other clear nights suffering from haze, moderate to
strong winds, or highly variable seeing.

The spectroscopic light curves were measured from the final,
calibrated spectra for each object by fitting a local, linear
continuum under the H$\beta$+[\ion{O}{3}] emission complex and
integrating the H$\beta$ emission-line flux above the fitted
continuum.  This technique includes the flux contribution from the
narrow H$\beta$ emission line, which is simply a constant offset in
the resultant light curves.  In the case of NGC\,5548, the red wing of
H$\beta$ extends underneath the [\ion{O}{3}] emission lines, so the
[\ion{O}{3}] lines were removed prior to measuring the H$\beta$ flux.
And for NGC\,6814, the continuum window to the blue of H$\beta$ had to
be placed to the blue of the \ion{He}{2} $\lambda 4686$ line as well
to avoid contamination from that emission line.
Table~\ref{table:fluxwind} gives the continuum windows and line
integration limits for each object as well as the mean and standard
deviation of the H$\beta$ flux.  We also list the mean continuum level
as the flux density at $5100 \times (1+z)$\,\AA.  The H$\beta$ light
curves for each of the objects are tabulated in
Tables~\ref{table:lc1}--\ref{table:lc3} and presented in
Figures~1--4 along with the $B$- and $V$-band
light curves.

Statistical properties of the H$\beta$ light curves are listed in
Table~\ref{table:variability} along with the properties of the $B$-
and $V$-band light curves for comparison.  Column (1) lists the
object, column (2) gives the measured feature, and column (3) lists
the number of measurements in each light curve.  For our analysis, we
binned all photometric measurements within 0.1\,days.  Columns (4) and
(5) are the sampling intervals between data points, measured as the
mean and median, respectively.  Column (6) gives the mean fractional
error, which is based on the comparison of observations that are
closely spaced in time.  Occasionally, spectra were obtained under
poor weather conditions and, in those cases, the uncertainties on the
H$\beta$ fluxes are given by photon counting statistics instead.  The
``excess variance'' in column (7) is computed as

\begin{equation}
F_{\rm var} = \frac{\sqrt{\sigma^2 - \delta^2}}{\langle f \rangle},
\end{equation}

\noindent where $\sigma^2$ is the variance of the fluxes, $\delta^2$ is 
their mean-square uncertainty, and $\langle f \rangle$ is the mean of
the observed fluxes.  Finally, column (8) is the ratio of the maximum to
the minimum flux ($R_{\rm max}$) for each light curve.

\section{Analysis}
\subsection{Time-Series Analysis}

\begin{figure*}
\plotone{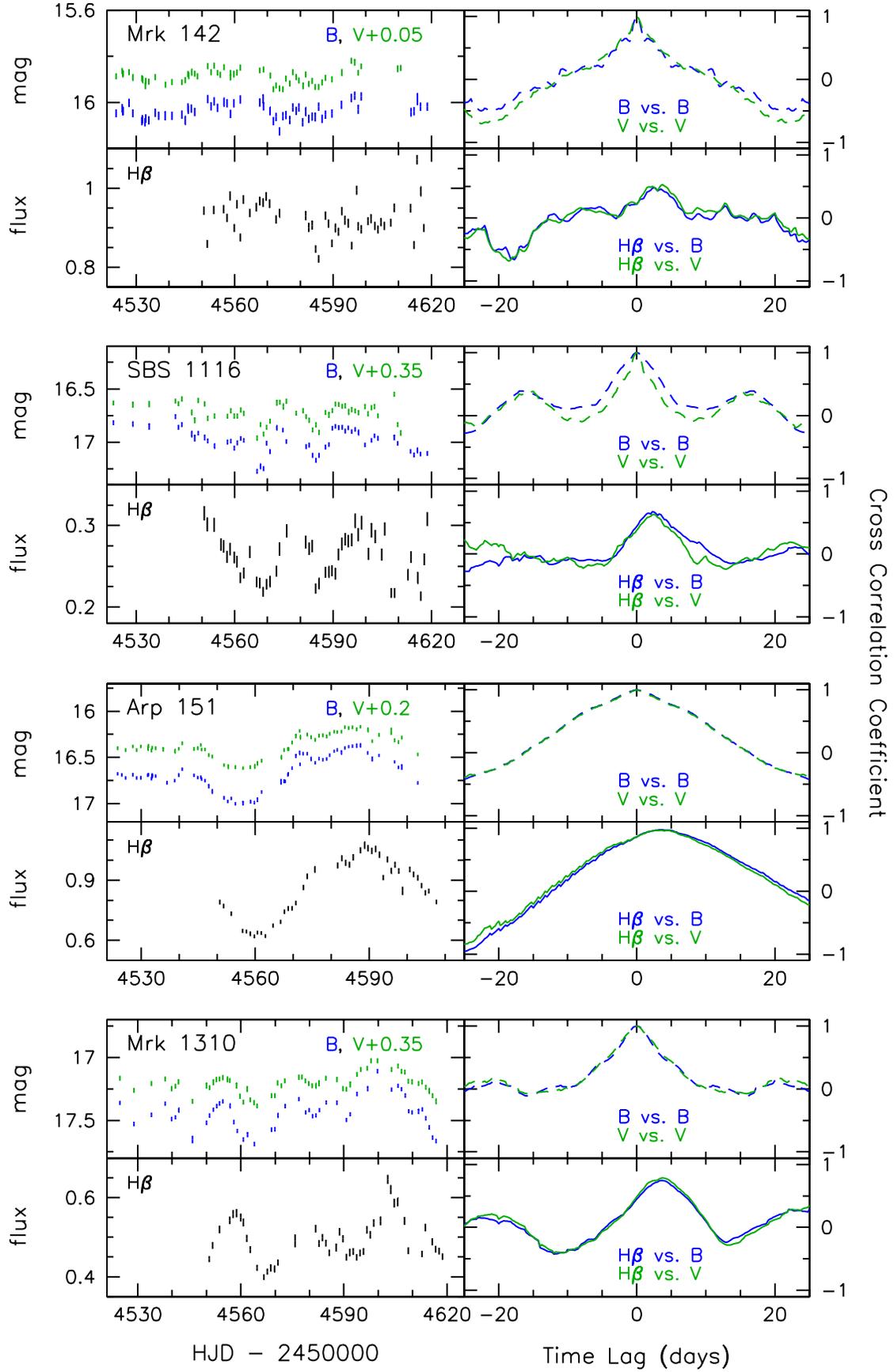}
\caption{{\it Left panels:} Photometric and H$\beta$ light curves for
         Mrk\,142, SBS\,1116+583A, Arp\,151, and Mrk\,1310.  The
         photometric measurements have units of Vega magnitudes, and
         the H$\beta$ emission-line fluxes have units of
         $10^{-13}$\,erg\,s$^{-1}$\,cm$^{-2}$. {\it Right panels:}
         Cross-correlation functions for the light curves.  For each
         object, the top panel shows the auto-correlation functions of
         the photometric light curves and the bottom panel shows the
         cross-correlation of H$\beta$ with the photometric light
         curves.}
\end{figure*}

\begin{figure*}
\plotone{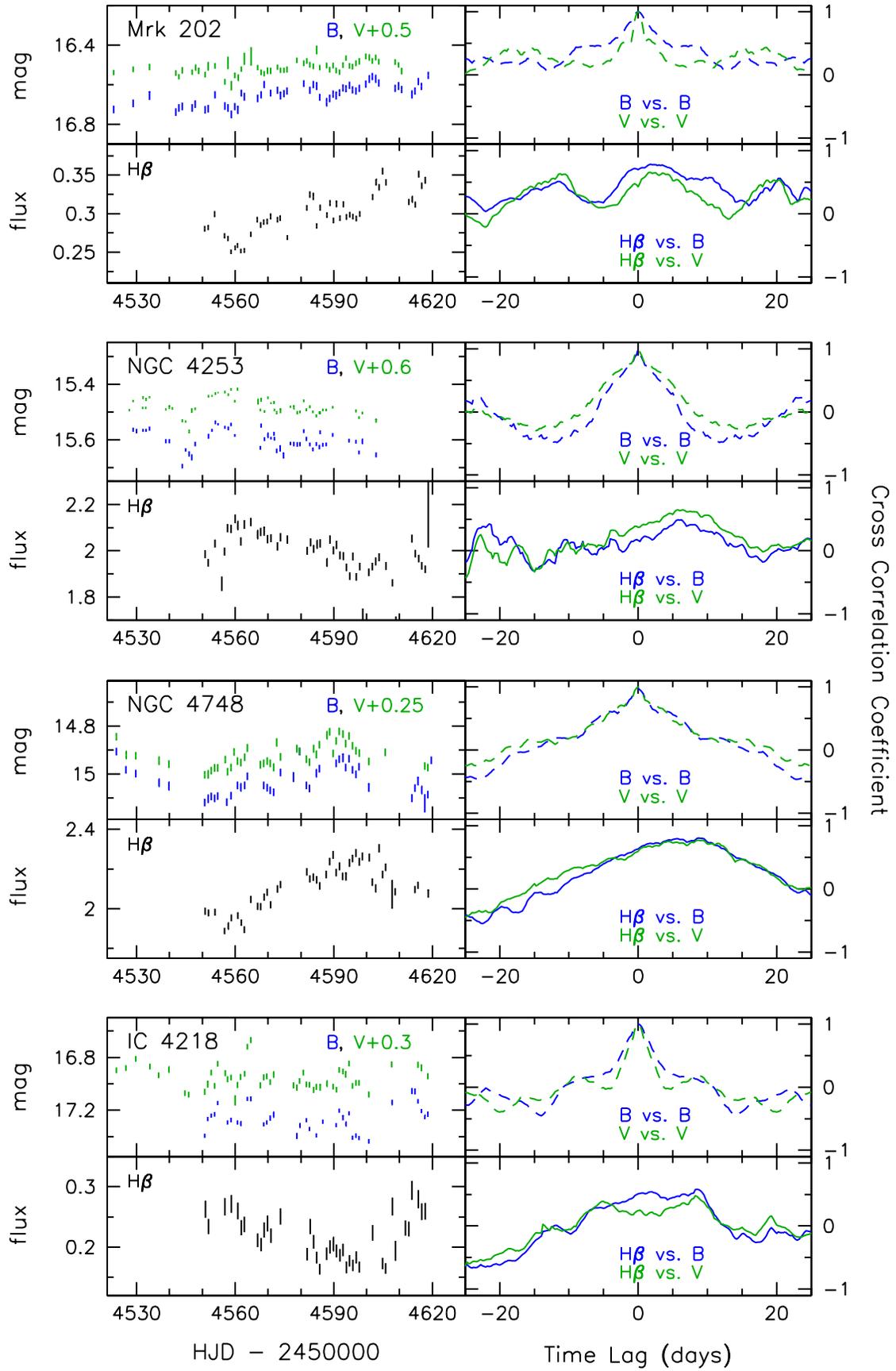}
\caption{Same as Figure 1 for Mrk\,202, NGC\,4253,
         NGC\,4748, and IC\,4218.}
\end{figure*}

\begin{figure*}
\plotone{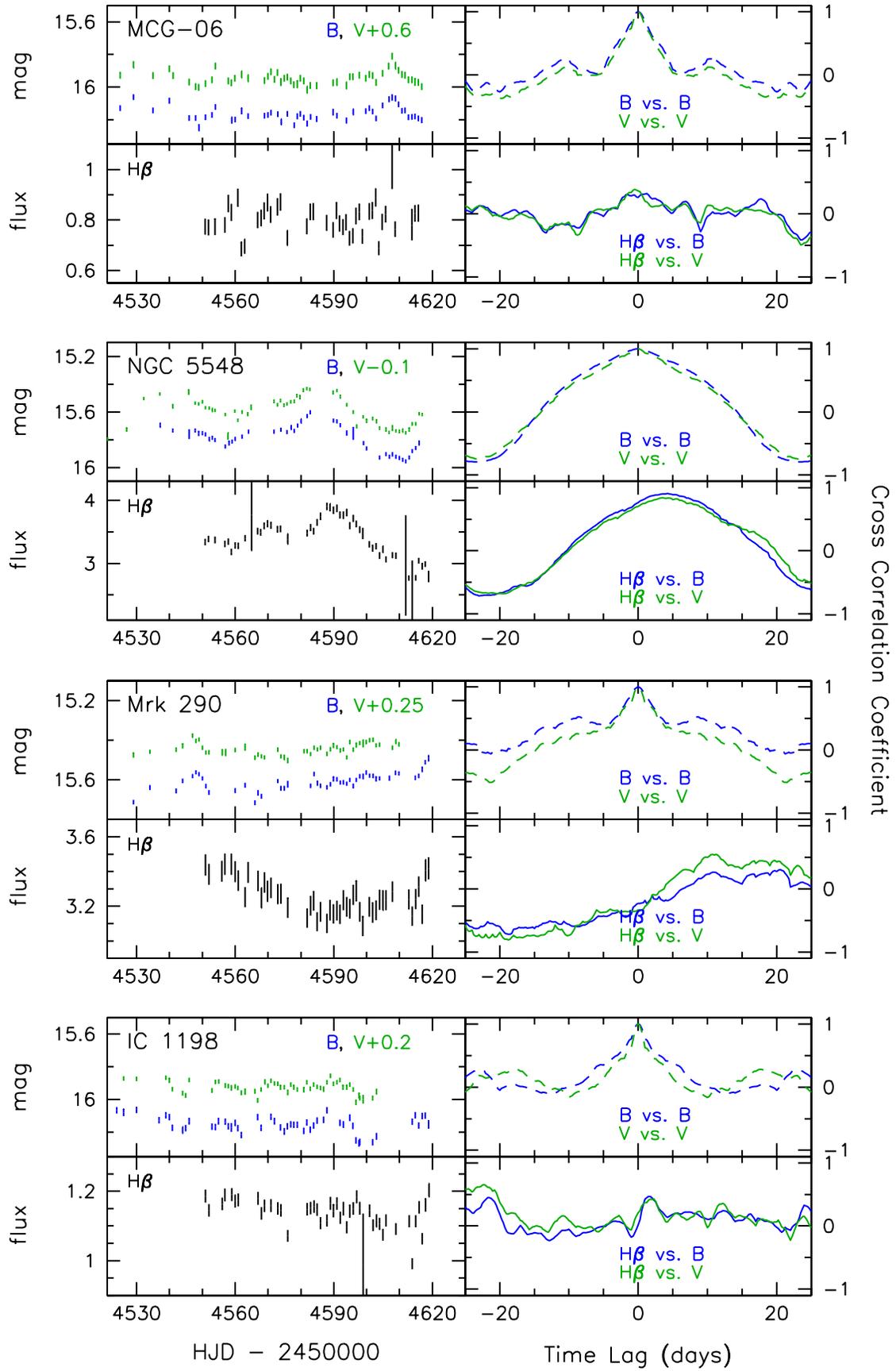}
\caption{Same as Figure 1 for MCG-06-30-15, NGC\,5548,
         Mrk\,290, and IC\,1198.}
\end{figure*}

\begin{figure*}
\plotone{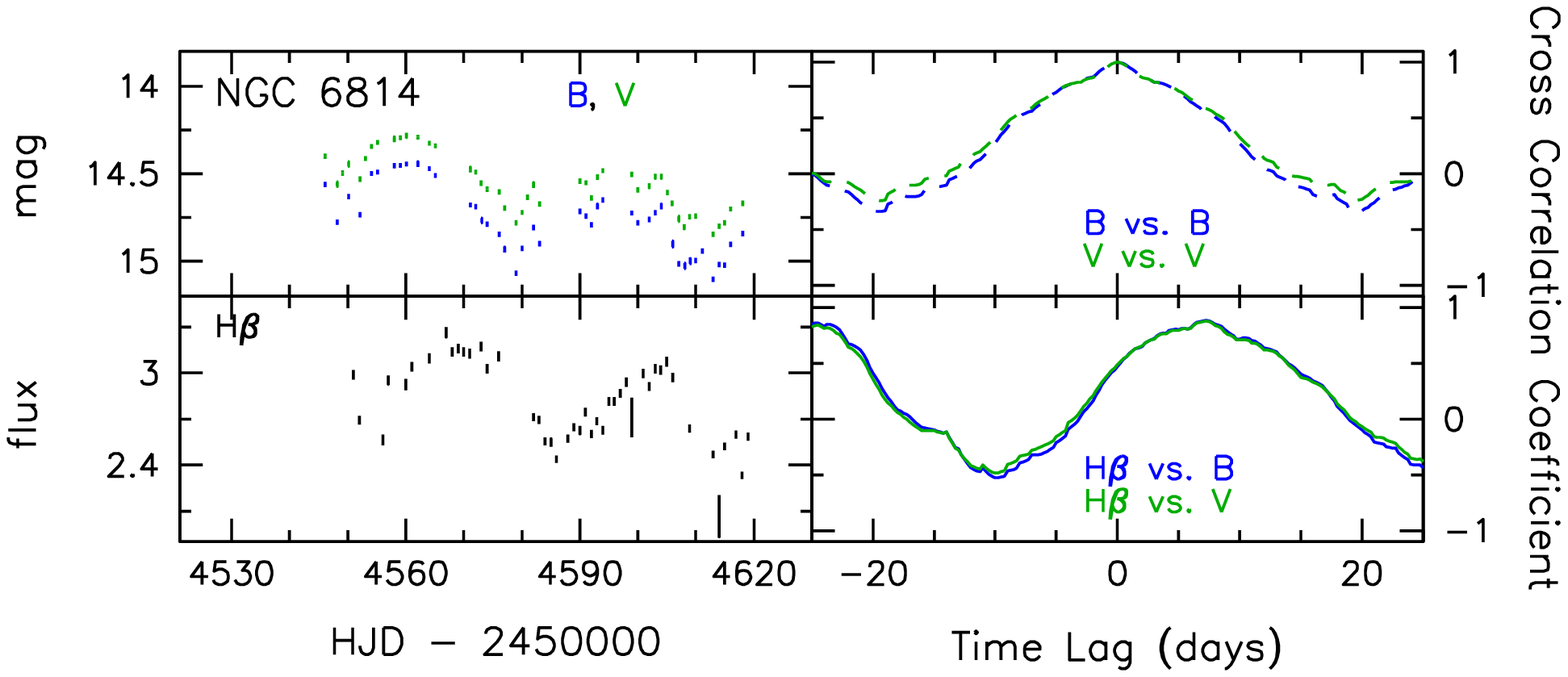}
\caption{Same as Figure 1 for NGC\,6814.}
\end{figure*}

For the time-series analysis, we consider both the $B$- and $V$-band
photometric light curves as the driving, continuum light curve.  In
general, they have similar sampling over the length of observations.
The variability in the $B$ band tends to be somewhat more pronounced
than in the $V$ band, most likely due to a smaller component of
host-galaxy starlight dilution, and is easily seen by comparing the
values of $F_{\rm var}$ and $R_{\rm max}$ for the $B$- and $V$-band
observations of each object as listed in
Table~\ref{table:variability}.  As shown in Paper~II, we find no
evidence for a time lag between the variations in the $B$ and $V$
bands.

To determine the average time lag between variations in the continuum
and variations in the H$\beta$ emission-line flux, we follow the
standard practice of cross-correlating the light curves.
Specifically, we employ the interpolation cross-correlation function
method (\citealt{gaskell86,gaskell87}) with the modifications
described by \citet{white94}. The method measures the
cross-correlation function between two light curves twice, first by
interpolating between the continuum points, and second by
interpolating between the emission-line points.  The average of the
two results is the final cross-correlation function (CCF).  Following
\citet{peterson04}, each CCF is characterized by the maximum
cross-correlation coefficient ($r_{\rm max}$), the time delay
corresponding to the location of $r_{\rm max}$ ($\tau_{\rm peak}$),
and the centroid of the points about the peak ($\tau_{\rm cent}$)
above some threshold value, typically $0.8r_{\rm max}$.

Figures~1--4 show the CCFs for the 13 AGNs in
our sample.  As mentioned above, we cross-correlated the H$\beta$ flux
with both the $B$- and $V$-band light curves, and we show the results
of both for comparison.  We also show the auto-correlation functions
for the photometric light curves, which, as expected, peak at a time
lag of zero days.  Four of the objects do not appear to show a
significant lag signal in their CCFs.  IC\,4218 has a broad,
flat-topped (H$\beta$ vs. $B$) or double-horned (H$\beta$ vs. $V$) CCF
structure centered around zero lag.  MCG-06-30-15 shows a noisy CCF
profile that appears to be consistent with zero at all lag times.
Mrk\,290 has a very slowly rising and flat-topped CCF profile at
positive lag times.  Inspection of the H$\beta$ variations in this
object does not seem to show an echo of the photometric variations,
and there is no H$\beta$ signal in the variable spectrum of Mrk\,290.
And IC\,1198 shows a CCF profile that is rather noisy and centered
about zero at all lag times, with the largest peak occurring at a lag
of $\sim -22$\,days.  There does not appear to be any signal from
H$\beta$ in the variable spectrum of this object either.

While it is quite simple to determine the lag time between two time
series by measuring either $\tau_{\rm peak}$ or $\tau_{\rm cent}$, it
is more difficult to quantify the uncertainty in the measured lag
time.  The standard procedure is to employ the Monte Carlo ``flux
randomization/random subset sampling'' method described by
\citet{peterson98b,peterson04}.  The method takes $N$ random and
independent samplings from the $N$ points available in the light
curves, regardless whether a datum has been sampled already.
The uncertainty for a point that is sampled $1 \leq n \leq N$ times is
reduced by a factor of $n^{1/2}$, and in general the fraction of
points that are not selected in any particular realization is $\sim
1/e$.  This ``random subset sampling'' helps to quantify the amount of
uncertainty in the lag time that arises based on the contribution from
individual points in the light curve.  The flux values in this
randomly sampled subset are then randomly altered by a Gaussian
deviation of the flux uncertainty.  This ``flux randomization''
accounts for the uncertainty in the measured flux values.  The CCF is
calculated for the sampled and modified light curves, and $r_{\rm
max}$, $\tau_{\rm cent}$, and $\tau_{\rm peak}$ are measured and
recorded.  The process is repeated for 1000 realizations, and
distributions of correlation measurements are built up.  The means of
the cross-correlation centroid distribution and the cross-correlation
peak distribution are taken to be $\tau_{\rm cent}$ and $\tau_{\rm
peak}$, respectively.  The uncertainties on $\tau_{\rm cent}$ and
$\tau_{\rm peak}$ are set such that 15.87\% of the realizations fall
above and 15.87\% fall below the range of uncertainties, corresponding
to $\pm 1\sigma$ for a Gaussian distribution.

Table~\ref{table:tau} lists the measured lag times and uncertainties
for the nine objects with significant H$\beta$ lag signatures in their
CCFs.  Also listed are the lag times and uncertainties after
correction for the time-dilation factor of $1+z$.

\subsection{Line-Width Measurement}

\begin{figure*}
\plotone{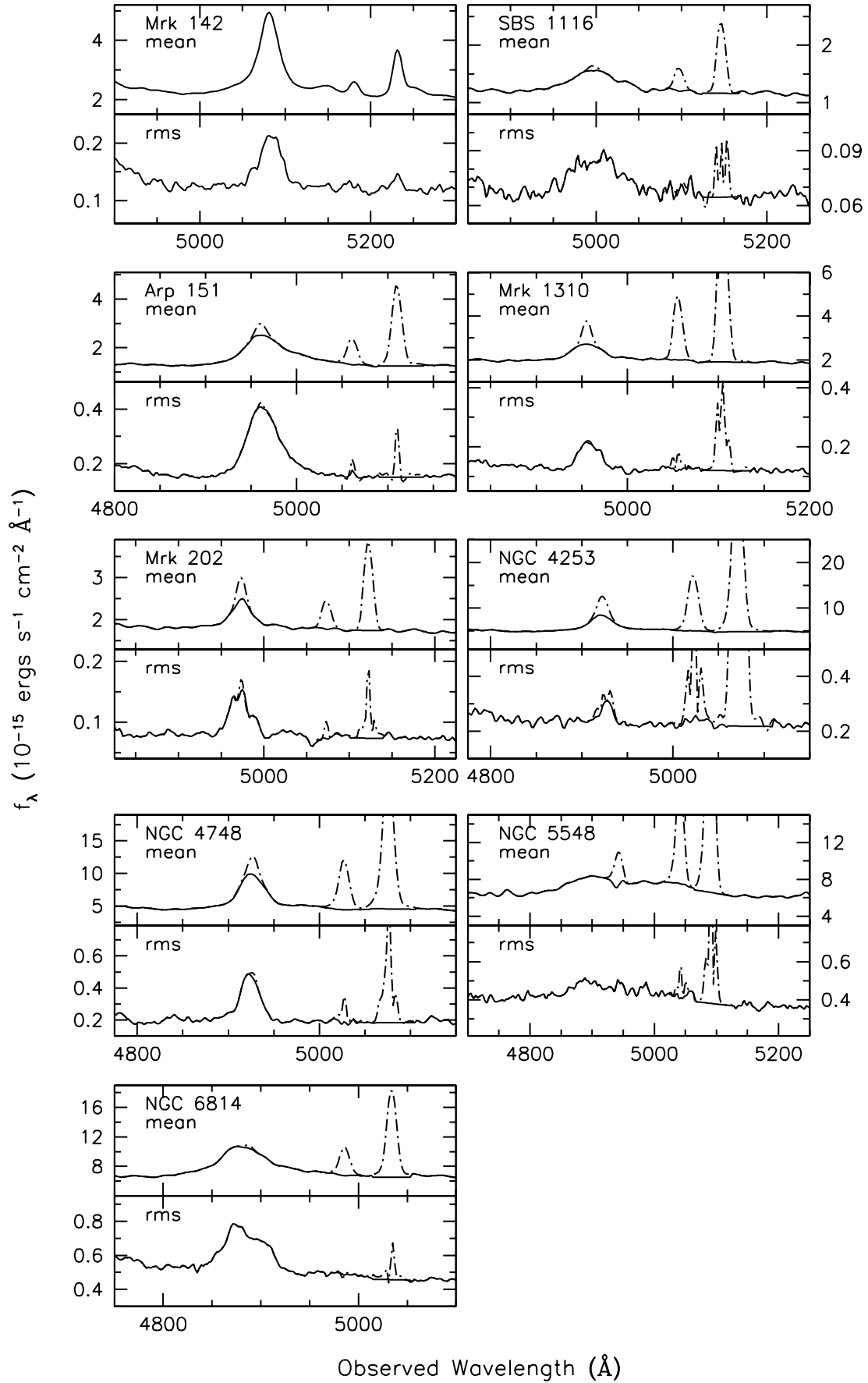}
\caption{Mean and variable (rms) spectra of the AGNs.  The solid lines
         are the narrow-line subtracted spectra, while the dot-dashed
         lines show the contributions from the H$\beta$ and
         [\ion{O}{3}] $\lambda\lambda$ 4959,5007 narrow lines.}
\end{figure*}

Figure~5 shows the mean and root-mean-square (rms) spectra in the
region around H$\beta$ for the nine objects with significant H$\beta$
lags.  For comparison, we include in Figure~6 the mean spectra of the
four objects with weak variability.  The rms spectra in Figure~5 show
the standard deviation of all the individual spectra relative to the
mean spectrum for an object, and are thus useful for visualizing and
quantifying the variable components of the spectra.  We also show the
narrow-line subtracted mean and rms spectra in Figure~5 (except for
Mrk\,142 which appears to have \ion{Fe}{2} emission blended with the
[\ion{O}{3}] emission in the mean spectrum).  We used the [\ion{O}{3}]
$\lambda 5007$ emission line as a template for the $\lambda 4959$ and
H$\beta$ narrow lines.  The ratio of [\ion{O}{3}] $\lambda
4959$/[\ion{O}{3}] $\lambda 5007$ was set at 0.34 \citep{storey00},
and Table~\ref{table:ratio} lists the derived ratios of
H$\beta$/[\ion{O}{3}] $\lambda 5007$.

The width of the broad H$\beta$ emission line was measured in the
narrow-line subtracted mean and rms spectra for each of the objects
and is reported as two separate measures: the full-width at
half-maximum flux (FWHM) and the line dispersion, $\sigma_{\rm line}$,
which is the second moment of the emission-line profile
\citep{peterson04}.  The uncertainties in the line widths are set
using a Monte Carlo random subset sampling method.  In this case, from
a set of $N$ spectra, a random subset of $N$ spectra is selected
without regard to whether a spectrum has previously been chosen and a
mean and rms spectrum are created from the chosen subset.  The FWHM
and $\sigma_{\rm line}$ are measured and recorded, and distributions
of line-width measurements are built up over 1000 realizations.  The
mean and standard deviation of each distribution are taken to be the
line width and uncertainty, respectively.
 
\begin{figure*}
\plotone{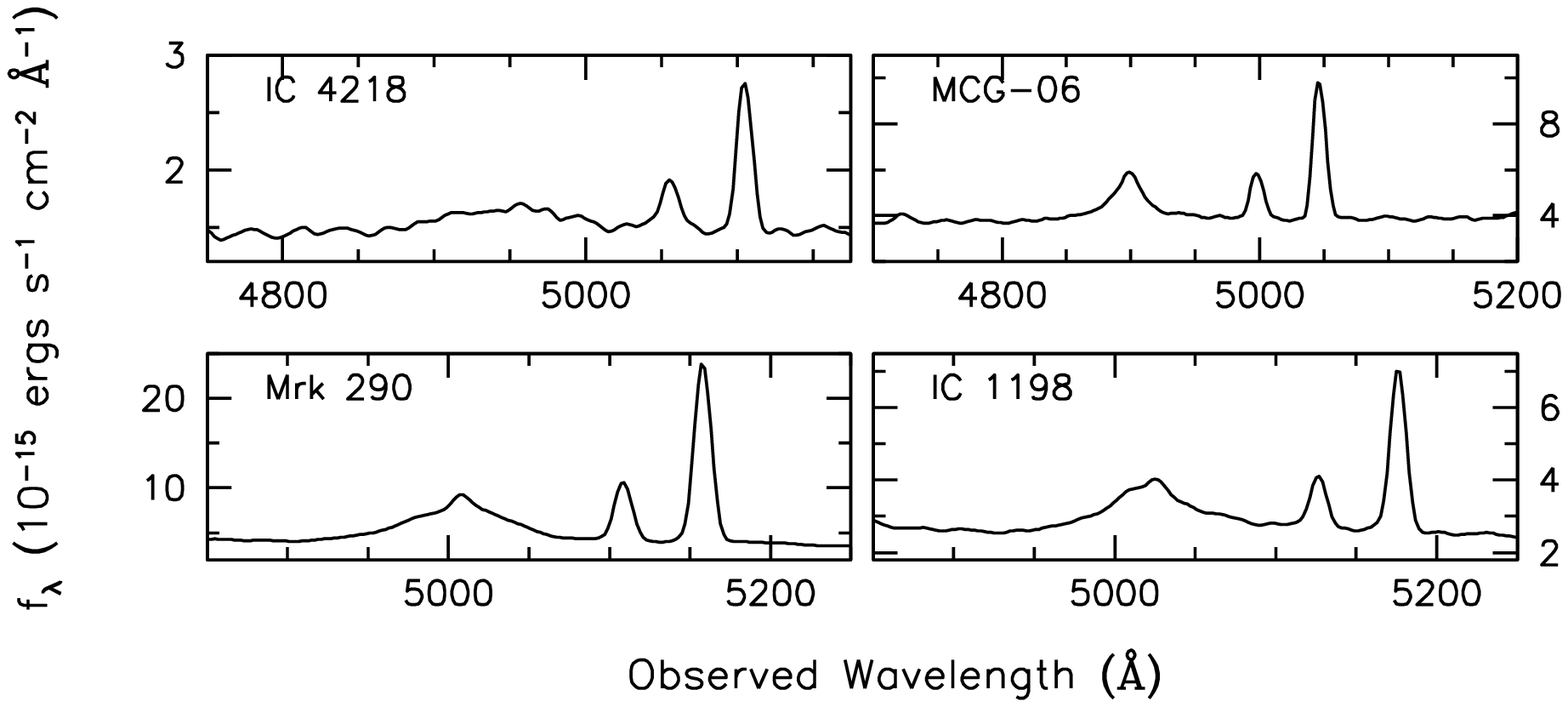}
\caption{Mean spectra of the AGNs without strong variability:
         IC\,4218, MCG$-06-30-15$, Mrk\,290, and IC\,1198.}
\end{figure*}

In a slight departure from the methods of \citet{peterson04}, we also
attempt to quantify the uncertainty from the exact placement of the
continuum.  For each object, we define a maximum continuum window
(typically 50\,\AA\ wide) on either side of the H$\beta$ +
[\ion{O}{3}] complex.  For each realization, a subset of the continuum
window on each side of at least 7 pixels (12\,\AA) is randomly
selected, from which the local linear continuum is fit.  In general we
find that this additional step does not affect the uncertainties of
the line widths measured in the rms spectra, but slightly increases
the errors from measurements made using the mean spectra.  This is not
particularly surprising, as the mean spectra have much higher
signal-to-noise ratios (S/N), so the exact placement of the continuum
window defines the specific low-level emission and absorption features
from the host-galaxy stellar population that will be included while
fitting the continuum.  These same low-level features are not detected
in an rms spectrum, and the errors are instead dominated by the
specifics of which spectra are included.

Finally, we correct the measured line widths for the dispersion of the
spectrograph following \citet{peterson04}.  The observed line width,
$\Delta \lambda_{\rm obs}$, can be described as a combination of the
intrinsic line width, $\Delta \lambda_{\rm true}$, and the
spectrograph resolution, $\Delta \lambda_{\rm res}$, such that

\begin{equation}
\Delta \lambda_{\rm obs}^2 \approx \Delta \lambda_{\rm true}^2 + \Delta \lambda_{\rm res}^2.
\end{equation}

\noindent We take our measurements of the FWHM of [\ion{O}{3}]
$\lambda 5007$ as $\Delta \lambda_{\rm obs}$.  Given our slit width of
4\arcsec\ and typical seeing of 2\arcsec\ throughout the campaign, the
target AGNs do not fill the entire width of the slit and so we do not
measure the resolution from sky lines or arc lamps.  Instead, we
assume that the high-resolution measurements of the widths of
[\ion{O}{3}] $\lambda 5007$ for several of the AGNs from
\citet{whittle92} are $\Delta \lambda_{\rm true}$ (listed here in
Table~\ref{table:o3width} after transformation to our adopted units
and the observed frame of the galaxy).\footnote{The spectroscopic
apertures employed in the observations quoted by \citet{whittle92} are
generally smaller than those employed here; however, narrow-band
[\ion{O}{3}] imaging of a subset of our sample by \citet{schmitt03}
shows that the vast majority of the [\ion{O}{3}] emission comes from a
fairly compact region of $\sim 1$\arcsec\ in width.}  We are then able
to deduce $\Delta \lambda_{\rm res}$, the FWHM resolution of the
spectra (also listed in Table~\ref{table:o3width}), with which
we are able to correct the measurements of the width of the broad
H$\beta$ line.  For those objects where measurements are not available
from \citeauthor{whittle92}, we assume a FWHM resolution of 12.5\,\AA,
which is within the range of measured spectral dispersions tabulated
in Table~\ref{table:o3width}, but slightly less than the median of
13.0\,\AA\ in an attempt to not overcorrect the velocity widths in
objects where we do not have a measurement of the intrinsic width of
the narrow lines.  The slight spread in measured dispersions is a
combination of factors including seeing, guiding, and the angular size
of the narrow-line region in each object.

We list the rest-frame, resolution-corrected broad H$\beta$ line width
measurements in Table~\ref{table:hbwidth}, from the mean and the rms
spectra of each of the nine objects with significant H$\beta$ lag
signatures.  The average ratio of H$\beta$ line widths measured in the
mean spectra to those in the rms spectra is $1.4 \pm 0.3$ for
$\sigma_{\rm line}$ and $1.3 \pm 0.7$ ($1.2 \pm 0.2$ excluding
NGC\,4748) for FWHM.  The average ratio of FWHM/$\sigma_{\rm line}$ is
$1.89 \pm 0.07$ in the mean spectra and $2.0 \pm 1.1$ ($2.1 \pm 0.5$
excluding NGC\,4748) in the rms spectra.  This is consistent with the
findings of \citet{collin06} that AGNs with narrow broad-line
components (i.e., $\sigma_{\rm line} < 2000$\,km\,s$^{-1}$) have
ratios of FWMH/$\sigma_{\rm line}$ that are less than the expected
value for a Gaussian line profile of 2.35.  It is also worth noting
that NGC\,5548, which currently has very broad H$\beta$ line widths
($\sigma_{\rm line} > 2000$\,km\,s$^{-1}$) has ratios of
FWMH/$\sigma_{\rm line} > 2.35$ in both the mean and rms spectra.

\subsection{Black Hole Mass}

Determination of black hole masses from reverberation mapping rests
upon the assumption that the gravity of the central, supermassive
black hole dominates the motions of the gas in the BLR.  The existence
of a ``virial'' relationship between time lag and line width, $v
\propto \tau^{-0.5}$, has been clearly shown in NGC\,5548
\citep{peterson99}, and has been in seen in numerous other objects
(\citealt{peterson00,onken02,kollatschny03}), upholding this basic
assumption.

The black hole mass is determined via the virial equation

\begin{equation}
    M_{\rm BH} = f \frac{c \tau v^2}{G},
\label{eqn:mbh}
\end{equation}

\noindent where $\tau$ is the mean time delay for the region of
interest (here, the H$\beta$-emitting region), $v$ is the velocity of
gas in that region, $c$ is the speed of light, $G$ is the
gravitational constant, and $f$ is a scaling factor of order unity
that depends on the detailed geometry and kinematics of the
line-emitting region.

\citet{peterson04} demonstrate that the combination of $\tau_{\rm
cent}$ and $\sigma_{\rm line, rms}$ provides the most robust
measurement of the black hole mass. By comparing the resultant masses
derived from several emission lines and independent datasets for the
same objects, the combination of $\tau_{\rm cent}$ and $\sigma_{\rm
line,rms}$ results in the least amount of scatter in the resultant
masses of all the combinations possible between the various line width
and lag time measures.  For the derived black hole masses presented
here, we will therefore adopt the combination of $\tau_{\rm cent}$ and
$\sigma_{\rm line,rms}$.

The absolute scaling of reverberation masses, the $f$ factor in
Equation~\ref{eqn:mbh}, is currently unknown.  Rather than assuming a
specific value of $f$ (e.g., \citealt{netzer90b}), and therefore
assuming specific physical details of the BLR, we instead adopt the
scaling factor determined by \citet{onken04} of $\langle f \rangle
\approx 5.5$.  This is the average value required to bring the $M_{\rm
BH} - \sigma_{\star}$ relationship for reverberation-mapped AGNs into
agreement with the $M_{\rm BH} - \sigma_{\star}$ relationship
determined for local, quiescent galaxies with dynamical mass
measurements.

Table~\ref{table:mbh} lists the black hole masses for the nine objects
presented in this work with H$\beta$ reverberation signals.  We list
both the ``virial product,'' which assumes that $f=1$, as well as the
adopted black hole mass using the \citet{onken04} scaling factor.
Figure~7 shows the range of black hole masses currently
probed by reverberation mapping.  The new masses determined here
(solid histogram, not including NGC\,5548) lie primarily in the range
$10^6$--$10^7$~M$_{\odot}$, in agreement with the expectations from
single-epoch estimates, and extending the range of black hole mass
coverage by a factor of $\sim 10$.

\begin{figure}
\plotone{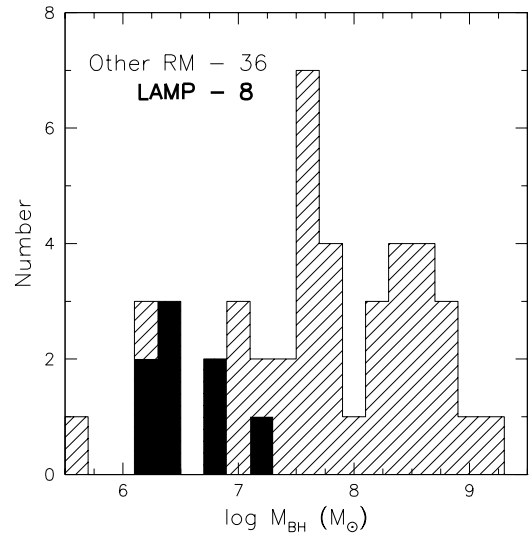}
\caption{Range of black hole masses currently probed by reverberation
         mapping experiments.  The 36 black hole masses that make up
         the hashed histogram come from \citet{peterson04,peterson05}
         and updates since then by
         \citet{bentz06b,denney06,bentz07,grier08}, and
         \citet{denney09}.  The eight new masses derived from the
         results presented here make up the solid histogram and
         primarily lie between $10^6$ and $10^7$~M$_{\odot}$}.
\end{figure}

\subsection{NGC\,5548: The Control Object}

\begin{figure}
\plotone{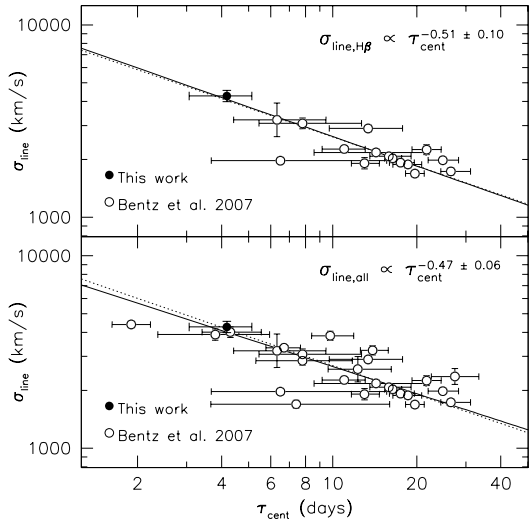}
\caption{Relationship between lag time and line width for several
         independent reverberation studies of NGC\,5548.  The top
         panel shows the relationship for H$\beta$ reverberation
         results only, while the bottom panel shows the relationship
         for all broad emission lines with reverberation results.  The
         dark circle in each panel is the H$\beta$ result from this
         work, while the open circles are the compilation of results
         from \citet{bentz07} and references therein.  The solid lines
         show the best fits to the relationship, with the slopes noted
         in each panel.  The dotted lines show the relationship with
         the slope fixed at the value expected for a virial
         relationship, i.e., $-0.5$.}
\end{figure}

\begin{figure}
\plotone{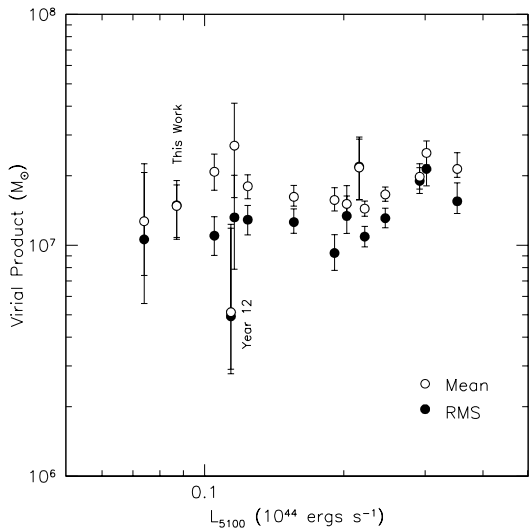}
\caption{Relationship between optical AGN luminosity and derived
         virial product for NGC\,5548.  The open circles show the
         virial product based on $\sigma_{\rm line}$ measured from the
         mean spectrum, and the filled circles are based on
         $\sigma_{\rm line}$ measured from the rms spectrum.}
\end{figure}

NGC\,5548 has by far the most independent reverberation-mapping
datasets of any individual AGN.  As a result, there is known to exist
a ``virial'' relationship between the broad-line width and the lag
time, which strongly suggests that the motions of the gas in the BLR
is dominated by a central supermassive object \citep{peterson99}.
Figure~8 shows this relationship for all of the independent H$\beta$
reverberation results for NGC\,5548, as well as the relationship for
all broad emission lines, including \ion{C}{4}, \ion{C}{3}], and
H$\alpha$.  The open circles are the results from previous
reverberation mapping campaigns, and the filled circle shows the
measurements of $\tau_{\rm cent}$ and $\sigma_{\rm line}$ for H$\beta$
presented here.  The H$\beta$ time lag presented here is the shortest
H$\beta$ lag measured for NGC\,5548, and is one of the shortest lags
measured for any emission line in NGC\,5548.  NGC\,5548 has been in a
very low luminosity state for the past several years (see
\citealt{bentz07}), and its current luminosity\footnote{The luminosity
at rest-frame 5100\,\AA\ has been corrected for the contribution from
starlight using {\it HST} imaging and the method of
\citet{bentz06a,bentz09b}.} of $\lambda L_{\lambda}({\rm 5100\AA})=
8.7 \times 10^{42}$\,erg\,s$^{-1}$ is only $\sim 20$\% brighter than
its lowest observed luminosity state in Spring 2005.  The low
luminosity of the AGN has resulted in a very broad, low-level,
double-peaked H$\beta$ emission-line profile in NGC\,5548, which does
increase the difficulty of accurately measuring the line width.
Despite this, the combination of lag time and line width measured here
falls where it is expected in Figure~8.

Additionally, we can compare the individual virial products for
NGC\,5548 as determined from each H$\beta$ reverberation dataset.
Figure~9 shows the virial product as a function of AGN
luminosity, with open circles representing the virial product based on
$\sigma_{\rm line}$ as measured in the mean spectrum, and filled
circles with $\sigma_{\rm line}$ from the rms spectrum.  While similar
to Figure~7 of \citet{bentz07}, the luminosities have been updated
with the new host-galaxy corrections of \citet{bentz09b}.  The point
denoted as ``Year 12'' is the monitoring dataset from the year 2000
and is known to be very poorly sampled and to yield ambiguous results
when the H$\beta$ light curve is cross-correlated with the continuum
light curve \citep{peterson02}.  The virial products from the time lag
and line widths presented here are consistent with previous results
within the observed scatter.  There does not seem to be any
significant trend over $\sim 0.6$\,dex in AGN luminosity, meaning that
the resultant virial product is not dependent on the luminosity state
of the AGN.

The agreement between the results for NGC\,5548 presented here and the
results from the previous 14 independent reverberation mapping
experiments for this same object shows that reverberation mapping is
both repeatable and reliable.  This agreement also shows that there
are no systematic biases in the LAMP analysis that would otherwise be
absent from similar high-quality reverberation mapping experiments.

\section{Velocity-Resolved Time Lags}

\begin{figure*}
\epsscale{1.1}
\plotone{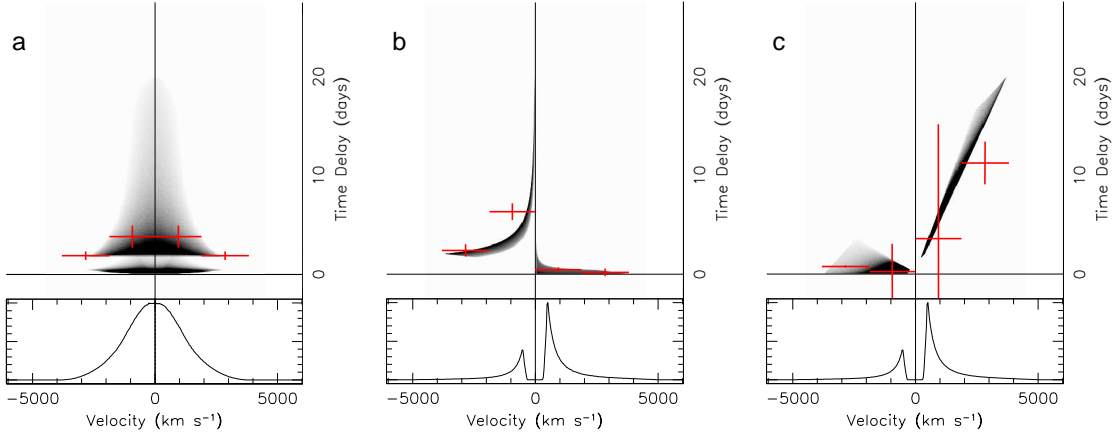}
\caption{Model transfer functions for broad-line regions with simple
         kinematics of (a) circular Keplerian orbits, (b)
         gravitational free-fall inflow, and (c) outflow with a
         constant acceleration (i.e., a Hubble or ballistic outflow).
         The gray-scale images show the full two-dimensional
         structure, while the vertical red error bars show the
         weighted mean and standard deviation of the time lag within
         discrete velocity bins that are represented by the horizontal
         red error bars.  For each of the three kinematic examples, the
         bottom panel shows the expected line profile (i.e., the
         two-dimensional structure integrated over all lag times).
         For each of the models, the line emission is restricted to a
         bicone with a semi-opening angle of 30\degr\ and the model is
         inclined at 20\degr\ so that the observer is inside the beam.
         The radiation structure within the BLR clouds is set so that
         the emission is enhanced for clouds at smaller radii, and the
         line emission is partially anisotropic, such that the
         emission is enhanced in the direction of the illuminating
         source.  The overall behavior of the red points is different
         for each of the three models: a symmetric structure around
         zero velocity for circular Keplerian orbits, longer lags in
         the blueshifted emission for infall, and longer lags in the
         redshifted emission for outflow. }
\end{figure*}

Up to this point, the discussion of the reverberation response for the
objects in the LAMP sample has centered around the average time lag
for the broad emission variability, which is related to the average
size of the H$\beta$-emitting BLR.  However, the average time lag is
simply the first moment of the so-called ``transfer function,'' which
describes the detailed line response as a function of time and
velocity (see \citealt{peterson01} for a full review).  

To illustrate a sample of possible expected transfer function
behaviors, Figure~10 shows model transfer functions for three
different kinematic states of the BLR: (a) circular Keplerian orbits,
(b) gravitational free-falling inflow, and (c) a constantly
accelerated outflow.  The BLR geometry and radiation parameters are
the same for each model: the emission is restricted to a biconical
structure with a semi-opening angle of 30\degr\ and an inclination of
20\degr, such that the observer is inside the beam.  The line emission
is enhanced for clouds at smaller radii, and is partially anisotropic
with enhanced radiation in the direction of the source.  Each
resulting model is a physically motivated and relatively plausible,
although likely simplified, model of an AGN BLR (for additional
models, see e.g., \citealt{welsh91,horne04}).  While the details of
the transfer function and emission line profile depend on the exact
geometry and line emission mechanics in the model, the overall
behavior for each kinematic state does not really change: BLR clouds
with circular orbits produce a symmetric response around zero
velocity, while inflow produces longer lag times in the blueshifted
emission and outflow produces the opposite, or longer lags in the
redshifted emission.  Therefore, recovery of the transfer function can
be an extremely powerful tool for discriminating between plausible
models for the BLR and is, in fact, the immediate goal of
reverberation mapping experiments.

However, achieving this goal is technically and observationally
challenging.  Several techniques have been developed in an attempt to
grapple with the technical difficulties, including the Maximum
Entropy Method (MEM; \citealt{horne94}), subtractively optimized local
averages \citep{pijpers94}, and regularized linear inversion
\citep{krolik95}.  Reverberation datasets are limited in sampling
duration and generally irregularly sampled, which, coupled with flux
uncertainties that are usually only a factor of a few smaller than the
flux variability amplitude, has placed severe limitations on past
attempts at transfer function recovery.  A partially recovered
transfer function for the \ion{C}{4}--\ion{He}{2} region of NGC\,4151
was hampered by extremely strong absorption in the \ion{C}{4} line
core, but perhaps shows some evidence for radial infall
\citep{ulrich96}.  \citet{kollatschny03} explored the behavior of
several optical emission lines in the spectrum of Mrk\,110 and found
possible indications for radial outflow.  Unfortunately, these and the
few other published attempts in the past have yielded notoriously
ambiguous results, a fact which is best illustrated by the analyses of
the {\it HST} \ion{C}{4} dataset for NGC\,5548 by several independent
groups.  Each of the studies concluded by favoring a different and
conflicting model of the \ion{C}{4} emitting gas in the BLR of
NGC\,5548: no radial motion \citep{wanders95}, some radial infall
\citep{done96}, and radial outflow (\citealt{chiang96,bottorff97}),
and all of these conclusions were based on analysis of {\it the same
data.}

Failure to achieve the goal of recovering a full, unambiguous transfer
function has led to more stringent observational requirements for
reverberation mapping experiments, including higher and more regular
sampling rates, longer sampling durations, and higher spectral
resolution and S/N requirements for each of the individual spectra
(e.g., \citealt{horne04}).  All of these requirements were carefully
considered while planning the LAMP observations, although past
difficulties and the relatively low luminosities of the target AGNs
did not immediately promote transfer function recovery as a main goal
of this project.  Because a full analysis of the reverberation data
presented here using the MEM or other techniques is beyond the scope
of this paper, we instead investigated whether there appeared to be
any strong signals of velocity-resolved time lag information in the
LAMP datasets.  For the six objects with the clearest average time lag
signatures, we measured the average lag time as a function of velocity
by creating light curves from the H$\beta$ emission flux in several
(typically four) equal variable-flux bins across the line
profile. Each of these light curves was then cross-correlated with the
$B$-band photometric light curve using the methods described in
Section~3.1.  We discuss the details for the six objects below.

\subsection{Individual Objects}

\paragraph{SBS\,1116+583A}

\begin{figure}
\plotone{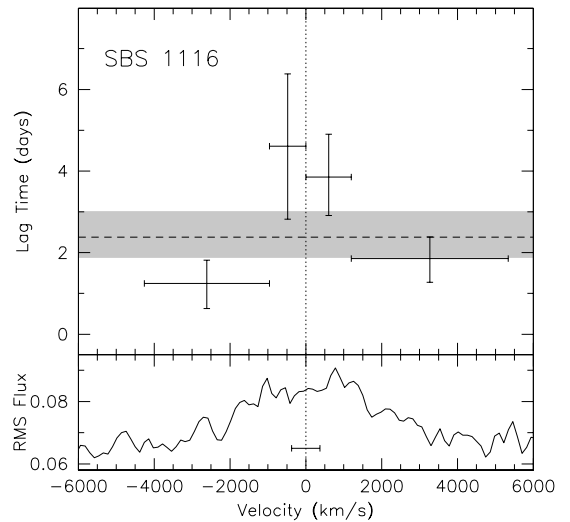}
\caption{Velocity-resolved time lag response (top panel) within the
         variable broad H$\beta$ emission (bottom panel) in
         SBS\,1116+583A.  In the top panel, the vertical error bars
         show the $1\sigma$ uncertainties on the time lag within each
         velocity bin denoted by the horizontal error bars.  The
         horizontal dashed line and gray band mark the average time
         lag and the $1\sigma$ uncertainty, respectively, for the
         entire emission line as listed in Table~\ref{table:tau}.  In
         the bottom panel, the horizontal error bar shows the FWHM
         velocity resolution determined in Section~3.2.}
\end{figure}

While the rms spectrum of this object is rather noisy, there is a
clear signature of H$\beta$ variability.  The H$\beta$ line was
divided up into four velocity bins, two on the blueshifted side and
two on the redshifted side, with each bin containing $\sim 1/4$ the
variable H$\beta$ flux.  Figure~11 shows the average lag time for each
of these bins as a function of bin velocity relative to the line
center.  The lag times in the wings of the emission line are not
consistent with the measured lag time in the line core, and the
profile shows a distinct, symmetric pattern around the line center, as
would be expected from a simple model of BLR gas in circular orbits.

\paragraph{Arp\,151}

\begin{figure}
\plotone{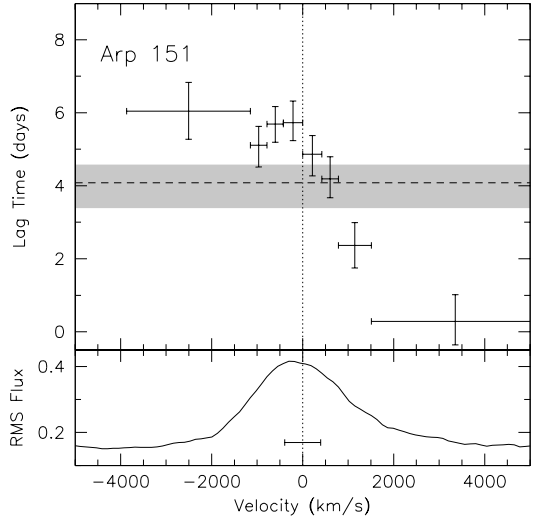}
\caption{Same as Figure~11 for Arp\,151.}
\end{figure}

A similar analysis for Arp\,151 was published in Paper~I, and here we
have updated the analysis to include the slight changes in the data
processing.  The result is that Figure~12 is not significantly
different from Figure~4 of Paper~I, and the lag time as a function of
velocity in the BLR of Arp\,151 shows a significantly asymmetric
profile, with longer lags in the blueshifted gas, and shorter lags in
the redshifted gas.  This pattern is consistent with the expectations
from a simple gravitational infall model.

\paragraph{Mrk\,1310}

\begin{figure}
\plotone{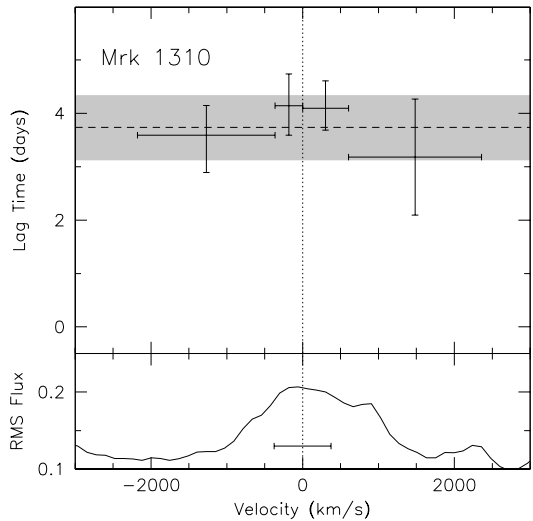}
\caption{Same as Figure~11 for Mrk\,1310.}
\end{figure}

In the case of Mrk\,1310, Figure~13 is rather ambiguous.  There is a
hint of slightly longer lag times in the line core; however, all of
the lags measured in the four velocity bins are consistent with a
single value, within the errors.  This particular structure is likely
consistent with circularly orbiting gas, as there does not seem to be
any evidence for a strong redward or blueward asymmetry that would
imply radial motion.

\paragraph{NGC\,4748}

\begin{figure}
\plotone{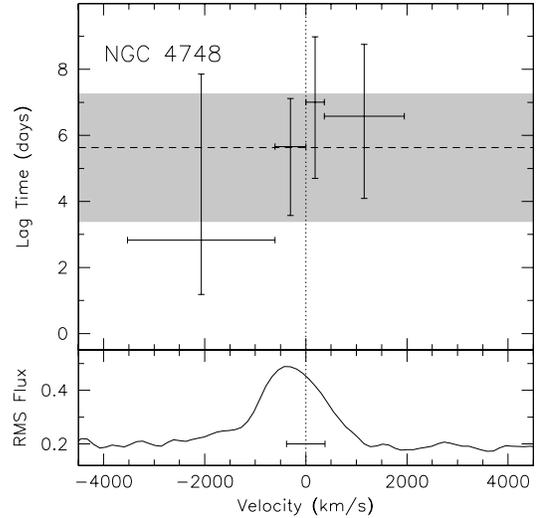}
\caption{Same as Figure~11 for NGC\,4748.}
\end{figure}

Examination of Figure~14 shows that there could be evidence for an
outflow in the BLR of NGC\,4748.  The extremely broad shape of the
cross-correlation functions for the H$\beta$ flux in this object (see
Figure~1) combined with the relatively low-level flux variations
results in rather large uncertainties for the measured lag times in
this object.  Each of the four velocity bins has a lag time that is
consistent within $\sim 1.5\sigma$ of the lag times measured for the
other bins, and so the significance of the velocity-resolved structure
for NGC\,4748 is not clear.

\paragraph{NGC\,5548}

\begin{figure}
\plotone{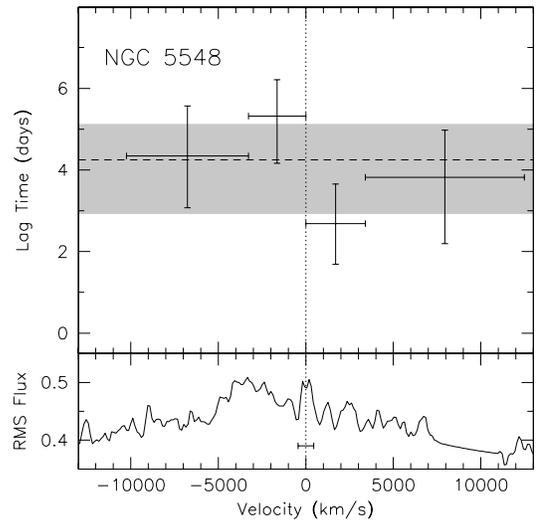}
\caption{Same as Figure~11 for NGC\,5548.}
\end{figure}

The current low-luminosity state of NGC\,5548 has resulted in a very
low, broad H$\beta$ line profile which extends under the [\ion{O}{3}]
doublet.  As the [\ion{O}{3}] lines in this object are quite strong,
we attempted to subtract them from each spectrum using a very
localized linear continuum (which actually includes the red wing of
the H$\beta$ profile) before creating the light curves for the four
velocity bins.  Only the most redward velocity bin is affected by the
[\ion{O}{3}] lines, and so the measured lag time for that bin may be
somewhat suspect.  The average lag time for each bin is shown in
Figure~15, where there does not seem to be an ordered behavior.  In
this object as well, each of the measured lag times is generally
consistent with the others within the errors, rendering interpretation
as somewhat ambiguous.

\paragraph{NGC\,6814}

\begin{figure}
\plotone{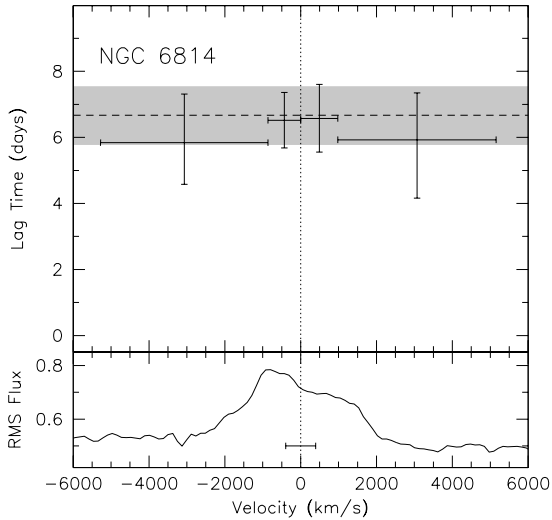}
\caption{Same as Figure~11 for NGC\,6814.}
\end{figure}

The lag structure for NGC\,6814 as a function of velocity is shown in
Figure~16, which again demonstrates that the lag time measured for
each velocity bin is consistent with a constant value, although there
is a slight preference for longer lag times in the line core than the
wings.  This behavior is most likely consistent with gas in circular
orbits.

\subsection{Discussion}

Although several of the objects examined here and presented in
Figures~11--16 show somewhat ambiguous or flat time lag behavior as a
function of velocity, both SBS\,1116 and Arp\,151 seem to show clear,
and yet completely different, behaviors.  The H$\beta$ response in
SBS\,1116 seems to be consistent with simple, circularly orbiting gas,
while gravitational infall seems to be the simplest picture for the
H$\beta$ response in Arp\,151.  The different behaviors of lag time as
a function of velocity for these two objects may be a clue that BLR
structure is very diverse from one object to another, even possibly an
evolutionary effect.  As such, SBS\,1116 and Arp\,151 are two
excellent targets for further and more detailed analysis using the MEM
or other techniques listed above.

While we plan to pursue recovery of full transfer functions for
SBS\,1116 and Arp\,151, we also plan to further examine the situation
for the other objects in our sample and determine whether the
perceived ambiguity in the velocity-resolved behavior is real or
merely a product of the simple analysis employed here.  Inspection of
the mean time lags in Figure~10a shows that the longer lags in the
emission-line core differ from the shorter lags in the wings by only
$\sim$1--2$\sigma$.  The addition of typical observational noise to
this model could conceivably alter the simplified behavior in
Figure~10a so that the red crosses are all consistent with a single
value, exactly as is seen for several of the objects here such as
Mrk\,1310 and NGC\,6814.

Recovery of a velocity-resolved transfer function for any of these
objects could place stringent limits on the $f$ factor in the
determination of the black hole mass for that particular object.
There is no reason to expect that the $f$ value is the same from
object to object, and differing $f$ values in individual objects may
be the main source of scatter in the AGN \msigma\ relationship (e.g.,
\citealt{collin06}).  The $\langle f \rangle$ of 5.5 employed in
Section~3.3 is empirically determined and does not assume any specific
details of the BLR geometry or kinematics, other than the dominance of
the black hole's gravity.  As this population average value has been
shown to remove any bias in the sample of reverberation masses when
compared to dynamical masses in quiescent galaxies, it is still
appropriate to use at this time, even though an individual object's
$f$ factor may differ.  We hope that further analysis of the
velocity-resolved information in the LAMP objects may begin to set
constraints on the $f$ factor for individual objects.

\section{Summary}

We have presented the H$\beta$ emission-line light curves and
reverberation analysis for the 13 AGNs included in the LAMP sample.
We measure H$\beta$ time lags relative to variations in the continuum
flux, which are related to the average sizes of the H$\beta$ BLRs, and
we derive black hole masses for the nine objects which display
significant time lag signatures.  In addition, we also explore the
velocity-resolved time lag behavior in 6 objects and find that the BLR
in SBS\,1116 seems to be consistent with a simple model of BLR gas in
circular orbits, while the BLR in Arp\,151 seems to be consistent with
gravitationally infalling gas.  More work is necessary to determine
what constraints may be set on the physical parameters of the BLR in
these two objects, as well as whether any constraints may be set for
other objects in the sample, although it seems clear that BLR
parameters may be very diverse among Type 1 AGNs.

Strong reverberation signals are also seen in other broad emission
lines for the objects in this sample, including H$\alpha$, H$\gamma$
and \ion{He}{2}, and future work will focus on the reverberation
signals in these emission lines.  We have a {\it Hubble Space
Telescope} Cycle 17 program (GO-11662, PI: Bentz) to image the host
galaxies of the AGNs in the LAMP sample, which will allow correction
for the host-galaxy starlight contribution to the continuum luminosity
for each object, and will allow us to extend the low-luminosity end of
the H$\beta$ \rl\ relationship, as well as the AGN \mlbulge\
relationship.  We also have new measurements of the bulge stellar
velocity dispersion for most of the objects in this sample, which will
allow us to extend the AGN \msigma\ relationship and explore any
updates to the population average $\langle f \rangle$ value in the
black hole mass determinations.  Finally, near-infrared photometric
monitoring data for a subset of the objects in this sample will allow
determination of the reverberation response of the dust torus in those
objects (e.g., \citealt{minezaki04,suganuma04}).

\acknowledgements

We would like to thank the excellent staff and support personnel at
Lick Observatory for their enormous help during our observing run, and
Brad Peterson for helpful conversations and the use of his analysis
software.  We also thank Josh Shiode for his observing help.  This
work was supported by NSF grants AST--0548198 (UC Irvine),
AST--0607485 (UC Berkeley), AST--0642621 (UC Santa Barbara), and
AST--0507450 (UC Riverside). The UC Berkeley researchers also
gratefully acknowledge the support of both the Sylvia \& Jim Katzman
Foundation and the TABASGO Foundation for the continued operation of
KAIT.  The work of DS was carried out at the Jet Propulsion
Laboratory, California Institute of Technology, under a contract with
NASA.

\clearpage


\begin{deluxetable}{lccccl}
\tablecolumns{6}
\tablewidth{0pt}
\tablecaption{Object List}
\tablehead{
\colhead{Object} &
\colhead{$\alpha_{2000}$} &
\colhead{$\delta_{2000}$} &
\colhead{$z$} &
\colhead{$A_{B}$\tablenotemark{a}} &
\colhead{Alternate} \\
\colhead{} &
\colhead{(hr min sec)} &
\colhead{($\degr\, \arcmin\, \arcsec$)} &
\colhead{} &
\colhead{(mag)} &
\colhead{Name}}
\startdata

Mrk\,142        & 10 25 31.3 & $+$51 40 35 & 0.04494 & 0.069 &  PG\,1022+519 \\
SBS\,1116+583A  & 11 18 57.7 & $+$58 03 24 & 0.02787 & 0.050 &  \\
Arp\,151        & 11 25 36.2 & $+$54 22 57 & 0.02109 & 0.059 &  Mrk\,40 \\
Mrk\,1310       & 12 01 14.3 & $-$03 40 41 & 0.01941 & 0.133 &  \\
Mrk\,202        & 12 17 55.0 & $+$58 39 35 & 0.02102 & 0.087 &  \\
NGC\,4253       & 12 18 26.5 & $+$29 48 46 & 0.01293 & 0.084 &  Mrk\,766 \\
NGC\,4748       & 12 52 12.4 & $-$13 24 53 & 0.01463 & 0.223 &  \\
IC\,4218        & 13 17 03.4 & $-$02 15 41 & 0.01933 & 0.132 &  \\
MCG-06-30-15    & 13 35 53.8 & $-$34 17 44 & 0.00775 & 0.266 &  ESO\,383-G035\\
NGC\,5548       & 14 17 59.5 & $+$25 08 12 & 0.01718 & 0.088 &  \\
Mrk\,290        & 15 35 52.3 & $+$57 54 09 & 0.02958 & 0.065 &  PG\,1543+580 \\
IC\,1198        & 16 08 36.4 & $+$12 19 51 & 0.03366 & 0.236 &  Mrk\,871 \\
NGC\,6814       & 19 42 40.6 & $-$10 19 25 & 0.00521 & 0.790 &  \\

\enddata
\label{table:objects}
\tablenotetext{a}{The Galactic extinction is based on \citet{schlegel98}.}
\end{deluxetable}


\begin{deluxetable}{lcccc}
\tablecolumns{5}
\tablewidth{0pt}
\tablecaption{Observation Log}
\tablehead{
\colhead{Object} &
\colhead{PA} &
\colhead{$t_{\rm exp}$} &
\colhead{S/N\tablenotemark{a}} &
\colhead{$\sec z$\tablenotemark{b}} \\
\colhead{} &
\colhead{($\degr$)} &
\colhead{(s)} &
\colhead{} &
\colhead{}} 

\startdata

Mrk\,142        &  90  & $2 \times 900$  & 90   &  1.06  \\
SBS\,1116+583A  &  90  & $2 \times 1200$ & 80	&  1.11  \\     
Arp\,151        &  90  & $2 \times 600$  & 80	&  1.10  \\
Mrk\,1310       &  90  & $2 \times 900$  & 60	&  1.34  \\
Mrk\,202        &  180 & $2 \times 900$  & 100	&  1.24  \\
NGC\,4253       &  60  & $2 \times 450$  & 120	&  1.30  \\
NGC\,4748       &  180 & $2 \times 450$  & 120	&  1.59  \\
IC\,4218        &  45  & $2 \times 900$  & 100	&  1.45  \\
MCG-06-30-15    &  180 & $2 \times 900$  & 80	&  3.15  \\
NGC\,5548       &  60  & $2 \times 300$  & 110	&  1.17  \\
Mrk\,290        &  90  & $2 \times 450$  & 110 	&  1.10  \\
IC\,1198        &  45  & $2 \times 1200$ & 160	&  1.17  \\
NGC\,6814       &  150 & $2 \times 900$  & 200	&  1.63  \\

\enddata
\tablenotetext{a}{The typical signal-to-noise per pixel in the continuum at $5100(1+z)$ \AA.}
\tablenotetext{b}{The median airmass at which the spectra were obtained throughout the campaign.}
\label{table:observ}
\end{deluxetable}


\begin{deluxetable}{lcll}
\tablecolumns{4}
\tablewidth{0pt}
\tablecaption{[\ion{O}{3}] $\lambda 5007$ Absolute Flux}
\tablehead{
\colhead{Object} &
\colhead{$f$([\ion{O}{3}])} &
\colhead{$f$([\ion{O}{3}])$_{\rm lit}$} &
\colhead{Ref.}\\ 
\colhead{} &
\colhead{($10^{-13}$ erg s$^{-1}$ cm$^{-2}$)} &
\colhead{($10^{-13}$ erg s$^{-1}$ cm$^{-2}$)} &
\colhead{}}
\startdata

Mrk\,142        & 0.321 & 0.20   & 1 \\
SBS\,1116+583A  & 0.158 &	 &   \\
Arp\,151       	& 0.489 & 0.73	 & 1 \\
Mrk\,1310      	& 1.10  &	 &   \\
Mrk\,202       	& 0.271 &	 &   \\
NGC\,4253      	& 5.52  & 4.54   & 2 \\
NGC\,4748      	& 3.50  & 3.65	 & 2 \\
IC\,4218       	& 0.181 &	 &   \\
MCG-06-30-15   	& 0.856 & 0.753, 1.14	        & 2, 3   \\
NGC\,5548      	& 5.55  & 5.49, 3.6, $5.58 \pm 0.27$ & 1, 2, 4 \\
Mrk\,290       	& 2.75  & 2.40,3.42             & 1, 5   \\
IC\,1198       	& 0.751 & 0.61, 0.70            & 2, 3   \\
NGC\,6814      	& 1.62  & 1.37,1.44,1.61        & 1, 3, 6 \\
			     
\enddata

\tablerefs{1. \citet{yee80},   
           2. \citet{degrijp92},
	   3. \citet{morris88},  
	   4. \citet{peterson91}, 
	   5. \citet{weedman72}, 
	   6. \citet{sekiguchi90}.
}	      
\label{table:o3flux}
\end{deluxetable}


\begin{deluxetable}{lccccc}
\tablecolumns{6}
\tablewidth{0pt}
\tabletypesize{\footnotesize}
\tablecaption{H$\beta$ Continuum Windows and Integration Limits}
\tablehead{
\colhead{Object} &
\multicolumn{2}{c}{Continuum Windows} &
\colhead{Line Limits} &
\colhead{$<f(H\beta)>$} &
\colhead{$<f_{\lambda}(5100\times(1+z))>$}\\
\colhead{} &
\colhead{(\AA)} &
\colhead{(\AA)} &
\colhead{(\AA)} &
\colhead{($10^{-13}$ erg s$^{-1}$ cm$^{-2}$)} &
\colhead{($10^{-15}$ erg s$^{-1}$ cm$^{-2}$ \AA$^{-1}$)}
}
\startdata

Mrk\,142        & $4960-5000$ & $5300-5350$ & $5045-5125$ & $0.928 \pm 0.080$ & $2.05 \pm 0.19$   \\
SBS\,1116+583A  & $4875-4925$ & $5200-5250$ & $4925-5055$ & $0.262 \pm 0.028$ & $1.088 \pm 0.067$ \\
Arp\,151        & $4850-4890$ & $5175-5250$ & $4900-5040$ & $0.86 \pm 0.15$   & $1.21 \pm 0.15$   \\
Mrk\,1310       & $4850-4900$ & $5150-5200$ & $4900-5010$ & $0.495 \pm 0.054$ & $1.87 \pm 0.12$   \\
Mrk\,202        & $4875-4925$ & $5150-5200$ & $4925-5025$ & $0.299 \pm 0.027$ & $1.698 \pm 0.070$ \\
NGC\,4253       & $4820-4860$ & $5150-5200$ & $4860-4975$ & $1.99 \pm 0.10$   & $4.59 \pm 0.26$	  \\
NGC\,4748       & $4600-4650$ & $5150-5200$ & $4850-5000$ & $2.11 \pm 0.11$   & $4.36 \pm 0.21$	  \\
IC\,4218        & $4800-4850$ & $5150-5200$ & $4850-5030$ & $0.217 \pm 0.037$ & $1.72 \pm 0.18$	  \\
MCG-06-30-15    & $4750-4800$ & $5150-5200$ & $4850-4940$ & $0.806 \pm 0.069$ & $4.33 \pm 0.59$   \\
NGC\,5548       & $4725-4775$ & $5150-5200$ & $4775-5150$ & $3.39 \pm 0.33$   & $6.12 \pm 0.38$	  \\
Mrk\,290        & $4850-4900$ & $5200-5250$ & $4900-5085$ & $3.254 \pm 0.099$ & $3.56 \pm 0.13$	  \\
IC\,1198        & $4900-4940$ & $5250-5300$ & $4940-5100$ & $1.135 \pm 0.045$ & $2.81 \pm 0.17$	  \\
NGC\,6814       & $4540-4590$ & $5100-5150$ & $4800-4970$ & $2.81 \pm 0.26$   & $6.47 \pm 0.50$   \\

\enddata

\tablecomments{The H$\beta$ fluxes above include the contribution from
               the narrow-line component, and the flux density at
               rest-frame 5100\,\AA\ includes the contribution from
               host-galaxy starlight.}

\label{table:fluxwind}
\end{deluxetable}


\begin{deluxetable}{cccccccccc}
\tablecolumns{10}
\tablewidth{0pt}
\tabletypesize{\tiny}
\tablecaption{H$\beta$ Light Curves --- Mrk\,142, SBS\,1116+583A, Arp\,151, Mrk\,1310, Mrk\,202}
\tablehead{
\multicolumn{2}{c}{Mrk\,142} &
\multicolumn{2}{c}{SBS\,1116+583A} &
\multicolumn{2}{c}{Arp\,151} &
\multicolumn{2}{c}{Mrk\,1310} &
\multicolumn{2}{c}{Mrk\,202}\\
\colhead{HJD} &
\colhead{$f$(H$\beta$)} &
\colhead{HJD} &
\colhead{$f$(H$\beta$)} &
\colhead{HJD} &
\colhead{$f$(H$\beta$)} &
\colhead{HJD} &
\colhead{$f$(H$\beta$)} &
\colhead{HJD} &
\colhead{$f$(H$\beta$)}}

\startdata

4550.6599 &  $0.943  \pm 0.011$  & 4550.6925 &  $0.3150 \pm  0.0088$  & 4550.7180  &  $0.791 \pm  0.012$  &  4550.7726  &  $0.4448 \pm  0.0081$ &   4550.7434  &  $0.2805 \pm  0.0035$ \\
4551.6560 &  $0.859  \pm 0.010$  & 4551.7189 &  $0.3013 \pm  0.0084$  & 4551.7478  &  $0.770 \pm  0.012$  &  4551.8100  &  $0.4793 \pm  0.0088$ &   4551.7693  &  $0.2822 \pm  0.0035$ \\
4553.6576 &  $0.945  \pm 0.011$  & 4553.7176 &  $0.3009 \pm  0.0084$  & 4553.7470  &  $0.733 \pm  0.011$  &  4553.8092  &  $0.5196 \pm  0.0095$ &   4553.7721  &  $0.2995 \pm  0.0037$ \\
4555.8322 &  $1.255  \pm 0.014$  & 4555.8587 &  $0.2767 \pm  0.0077$  & 4556.7118  &  $0.647 \pm  0.010$  &  4556.7724  &  $0.5427 \pm  0.0099$ &   4556.7381  &  $0.2711 \pm  0.0034$ \\
4556.6591 &  $0.947  \pm 0.011$  & 4556.6866 &  $0.2743 \pm  0.0077$  & 4557.7555  &  $0.644 \pm  0.010$  &  4557.8255  &  $0.5583 \pm  0.0102$ &   4557.7884  &  $0.2677 \pm  0.0034$ \\
4557.6574 &  $0.925  \pm 0.010$  & 4557.6847 &  $0.2615 \pm  0.0073$  & 4558.6902  &  $0.632 \pm  0.009$  &  4558.7907  &  $0.5608 \pm  0.0103$ &   4558.7121  &  $0.2506 \pm  0.0031$ \\
4558.6464 &  $0.981  \pm 0.011$  & 4558.6633 &  $0.2710 \pm  0.0076$  & 4559.8700  &  $0.619 \pm  0.009$  &  4559.9485  &  $0.5503 \pm  0.0101$ &   4559.9658  &  $0.2556 \pm  0.0032$ \\
4559.8879 &  $0.899  \pm 0.010$  & 4559.9038 &  $0.2607 \pm  0.0073$  & 4560.6570  &  $0.638 \pm  0.010$  &  4560.7912  &  $0.5377 \pm  0.0098$ &   4560.7389  &  $0.2587 \pm  0.0032$ \\
4560.6809 &  $0.960  \pm 0.011$  & 4560.7082 &  $0.2556 \pm  0.0071$  & 4561.6791  &  $0.637 \pm  0.010$  &  4561.8143  &  $0.4968 \pm  0.0091$ &   4561.7450  &  $0.2516 \pm  0.0031$ \\
4561.7065 &  $0.875  \pm 0.010$  & 4561.7268 &  $0.2358 \pm  0.0066$  & 4562.7017  &  $0.622 \pm  0.009$  &  4562.7961  &  $0.4686 \pm  0.0086$ &   4562.7600  &  $0.2522 \pm  0.0032$ \\
4562.7222 &  $0.972  \pm 0.011$  & 4562.7373 &  $0.2419 \pm  0.0068$  & 4564.7136  &  $0.673 \pm  0.010$  &  4564.7669  &  $0.4225 \pm  0.0077$ &   4564.7344  &  $0.2731 \pm  0.0034$ \\
4564.6609 &  $0.937  \pm 0.011$  & 4564.6890 &  $0.2674 \pm  0.0075$  & 4566.7181  &  $0.692 \pm  0.010$  &  4566.7677  &  $0.3993 \pm  0.0073$ &   4566.7384  &  $0.2922 \pm  0.0037$ \\
4566.6662 &  $0.952  \pm 0.011$  & 4566.6937 &  $0.2307 \pm  0.0064$  & 4567.7170  &  $0.743 \pm  0.011$  &  4567.7764  &  $0.4129 \pm  0.0076$ &   4567.7396  &  $0.2855 \pm  0.0036$ \\
4567.6650 &  $0.968  \pm 0.011$  & 4567.6924 &  $0.2293 \pm  0.0064$  & 4568.7177  &  $0.759 \pm  0.011$  &  4568.7689  &  $0.4213 \pm  0.0077$ &   4568.7380  &  $0.2891 \pm  0.0036$ \\
4568.6636 &  $0.963  \pm 0.011$  & 4568.6919 &  $0.2187 \pm  0.0061$  & 4569.7258  &  $0.760 \pm  0.011$  &  4569.7798  &  $0.4200 \pm  0.0077$ &   4569.7496  &  $0.2899 \pm  0.0036$ \\
4569.6805 &  $0.979  \pm 0.011$  & 4569.7050 &  $0.2280 \pm  0.0064$  & 4570.7526  &  $0.786 \pm  0.012$  &  4570.8030  &  $0.4363 \pm  0.0080$ &   4570.7727  &  $0.2965 \pm  0.0037$ \\
4570.6594 &  $0.961  \pm 0.011$  & 4570.7310 &  $0.2312 \pm  0.0065$  & 4572.7551  &  $0.864 \pm  0.013$  &  4575.7581  &  $0.4908 \pm  0.0162$ &   4572.8944  &  $0.2922 \pm  0.0037$ \\
4572.6842 &  $0.921  \pm 0.010$  & 4572.7106 &  $0.2439 \pm  0.0068$  & 4573.7215  &  $0.919 \pm  0.014$  &  4581.7875  &  $0.5216 \pm  0.0096$ &   4573.7419  &  $0.2945 \pm  0.0037$ \\
4573.6698 &  $0.937  \pm 0.011$  & 4573.6979 &  $0.2705 \pm  0.0076$  & 4575.6875  &  $0.954 \pm  0.014$  &  4582.7344  &  $0.4841 \pm  0.0089$ &   4575.8692  &  $0.2691 \pm  0.0034$ \\
4575.7090 &  $0.637  \pm 0.007$  & 4575.9200 &  $0.2935 \pm  0.0082$  & 4581.7317  &  $0.974 \pm  0.015$  &  4583.7331  &  $0.4987 \pm  0.0091$ &   4581.7533  &  $0.3076 \pm  0.0039$ \\
4581.6673 &  $0.929  \pm 0.010$  & 4581.7033 &  $0.2796 \pm  0.0078$  & 4582.8300  &  $1.010 \pm  0.015$  &  4584.7524  &  $0.4851 \pm  0.0089$ &   4582.7752  &  $0.3249 \pm  0.0041$ \\
4582.6672 &  $0.900  \pm 0.010$  & 4582.6997 &  $0.2714 \pm  0.0076$  & 4583.8341  &  $0.987 \pm  0.015$  &  4585.7278  &  $0.4607 \pm  0.0084$ &   4583.7754  &  $0.3218 \pm  0.0040$ \\
4583.6648 &  $0.905  \pm 0.010$  & 4583.6927 &  $0.2750 \pm  0.0077$  & 4584.7895  &  $0.977 \pm  0.015$  &  4587.7413  &  $0.4731 \pm  0.0087$ &   4584.8230  &  $0.2843 \pm  0.0036$ \\
4584.6796 &  $0.846  \pm 0.010$  & 4584.7176 &  $0.2226 \pm  0.0062$  & 4585.8605  &  $1.014 \pm  0.015$  &  4588.7303  &  $0.5137 \pm  0.0094$ &   4585.9011  &  $0.3079 \pm  0.0039$ \\
4585.6626 &  $0.821  \pm 0.009$  & 4585.6924 &  $0.2288 \pm  0.0064$  & 4587.7764  &  $1.049 \pm  0.016$  &  4589.7353  &  $0.4965 \pm  0.0091$ &   4587.9043  &  $0.2983 \pm  0.0037$ \\
4587.6776 &  $0.931  \pm 0.010$  & 4587.7063 &  $0.2426 \pm  0.0068$  & 4588.7821  &  $1.074 \pm  0.016$  &  4590.7365  &  $0.4496 \pm  0.0082$ &   4588.8947  &  $0.3137 \pm  0.0039$ \\
4588.6690 &  $0.910  \pm 0.010$  & 4588.6966 &  $0.2431 \pm  0.0068$  & 4589.7693  &  $1.060 \pm  0.016$  &  4591.7273  &  $0.4599 \pm  0.0084$ &   4589.9091  &  $0.2915 \pm  0.0036$ \\
4589.6708 &  $0.861  \pm 0.010$  & 4589.7006 &  $0.2445 \pm  0.0068$  & 4590.7677  &  $1.046 \pm  0.016$  &  4592.7303  &  $0.4641 \pm  0.0085$ &   4590.9006  &  $0.3121 \pm  0.0039$ \\
4590.6762 &  $0.900  \pm 0.010$  & 4590.7037 &  $0.2401 \pm  0.0067$  & 4591.7584  &  $1.060 \pm  0.016$  &  4593.7325  &  $0.4493 \pm  0.0082$ &   4591.9069  &  $0.2942 \pm  0.0037$ \\
4591.6675 &  $0.890  \pm 0.010$  & 4591.6948 &  $0.2645 \pm  0.0074$  & 4592.7618  &  $1.027 \pm  0.015$  &  4594.7361  &  $0.4602 \pm  0.0084$ &   4592.9010  &  $0.3124 \pm  0.0039$ \\
4592.6697 &  $0.942  \pm 0.011$  & 4592.6977 &  $0.2773 \pm  0.0077$  & 4593.7636  &  $0.913 \pm  0.014$  &  4595.7465  &  $0.4639 \pm  0.0085$ &   4593.9007  &  $0.2960 \pm  0.0037$ \\
4593.6707 &  $0.918  \pm 0.010$  & 4593.6993 &  $0.2831 \pm  0.0079$  & 4594.8047  &  $1.003 \pm  0.015$  &  4596.7329  &  $0.5186 \pm  0.0095$ &   4594.8703  &  $0.2987 \pm  0.0037$ \\
4594.6761 &  $0.910  \pm 0.010$  & 4594.7031 &  $0.2809 \pm  0.0078$  & 4595.8015  &  $0.948 \pm  0.014$  &  4597.7353  &  $0.5064 \pm  0.0093$ &   4595.8579  &  $0.2964 \pm  0.0037$ \\
4595.6869 &  $0.855  \pm 0.010$  & 4595.7145 &  $0.2823 \pm  0.0079$  & 4596.8003  &  $0.978 \pm  0.015$  &  4598.7316  &  $0.5358 \pm  0.0098$ &   4596.8908  &  $0.2938 \pm  0.0037$ \\
4597.1746 &  $0.995  \pm 0.011$  & 4596.7005 &  $0.3058 \pm  0.0085$  & 4597.8005  &  $0.945 \pm  0.014$  &  4600.7030  &  $0.5400 \pm  0.0099$ &   4597.8964  &  $0.2997 \pm  0.0038$ \\
4597.6761 &  $0.904  \pm 0.010$  & 4597.7035 &  $0.2888 \pm  0.0081$  & 4598.8010  &  $0.846 \pm  0.021$  &  4602.7830  &  $0.6463 \pm  0.0118$ &   4601.8116  &  $0.3216 \pm  0.0040$ \\
4598.6721 &  $0.896  \pm 0.010$  & 4598.6992 &  $0.3034 \pm  0.0085$  & 4600.7205  &  $0.954 \pm  0.014$  &  4603.7310  &  $0.6206 \pm  0.0114$ &   4602.9024  &  $0.3408 \pm  0.0043$ \\
4600.6732 &  $0.917  \pm 0.010$  & 4600.8294 &  $0.2653 \pm  0.0074$  & 4601.7713  &  $0.928 \pm  0.014$  &  4604.7232  &  $0.5865 \pm  0.0107$ &   4603.8855  &  $0.3343 \pm  0.0042$ \\
4601.6753 &  $0.922  \pm 0.010$  & 4601.7108 &  $0.2936 \pm  0.0082$  & 4602.8002  &  $0.912 \pm  0.014$  &  4605.7092  &  $0.5888 \pm  0.0108$ &   4604.9140  &  $0.3554 \pm  0.0044$ \\
4602.8195 &  $0.905  \pm 0.010$  & 4602.8731 &  $0.2864 \pm  0.0080$  & 4603.8428  &  $0.873 \pm  0.013$  &  4607.7142  &  $0.5411 \pm  0.0099$ &   4605.8939  &  $0.3407 \pm  0.0043$ \\
4603.6917 &  $0.896  \pm 0.010$  & 4604.8290 &  $0.2642 \pm  0.0074$  & 4604.8036  &  $0.837 \pm  0.013$  &  4608.7901  &  $0.4663 \pm  0.0085$ &   4612.8862  &  $0.3151 \pm  0.0039$ \\
4604.6891 &  $0.914  \pm 0.010$  & 4605.8064 &  $0.2957 \pm  0.0083$  & 4605.7268  &  $0.846 \pm  0.013$  &  4612.7931  &  $0.4576 \pm  0.0084$ &   4613.8602  &  $0.3185 \pm  0.0040$ \\
4605.6784 &  $0.905  \pm 0.010$  & 4607.8182 &  $0.2171 \pm  0.0061$  & 4607.7301  &  $0.791 \pm  0.012$  &  4613.7813  &  $0.5243 \pm  0.0096$ &   4614.8871  &  $0.3116 \pm  0.0039$ \\
4607.6841 &  $0.913  \pm 0.010$  & 4608.8160 &  $0.2170 \pm  0.0061$  &	           &  	  	          &  4615.7819  &  $0.4884 \pm  0.0089$ &   4615.8847  &  $0.3514 \pm  0.0044$ \\
4608.6829 &  $0.950  \pm 0.011$  & 4612.8142 &  $0.2396 \pm  0.0067$  &	           &  	  	          &  4616.7337  &  $0.4694 \pm  0.0086$ &   4616.8548  &  $0.3364 \pm  0.0042$ \\
4613.6800 &  $0.949  \pm 0.011$  & 4613.8024 &  $0.2611 \pm  0.0073$  &	           &  	  	          &  4617.7604  &  $0.4647 \pm  0.0085$ &   4617.8826  &  $0.3433 \pm  0.0043$ \\
4614.6846 &  $0.856  \pm 0.010$  & 4615.8328 &  $0.2356 \pm  0.0066$  &	           &  	  	          &  4618.7642  &  $0.4511 \pm  0.0083$ &              &                       \\
4615.6835 &  $1.072  \pm 0.012$  & 4616.7816 &  $0.2132 \pm  0.0060$  &	           &  	  	          &             &                       &              &                       \\
4616.7547 &  $0.992  \pm 0.011$  & 4617.8077 &  $0.2587 \pm  0.0072$  &	           &  	  	          &             &                       &              &                       \\
4617.6966 &  $0.899  \pm 0.010$  & 4618.7821 &  $0.3079 \pm  0.0086$  &	           &  	  	          &             &                       &              &                       \\
4618.6954 &  $1.114  \pm 0.013$  & 	     &	    	              &	           &                      &             &                       &              &                       \\
		      		     
\enddata

\tablecomments{HJD = Heliocentric Julian Day$-$2450000; H$\beta$
               emission-line fluxes are in units of $10^{-13}$ erg
               s$^{-1}$ cm$^{-2}$.}
\label{table:lc1}
\end{deluxetable}


\begin{deluxetable}{cccccccccc}
\tablecolumns{10}
\tablewidth{0pt}
\tabletypesize{\tiny}
\tablecaption{H$\beta$ Light Curves --- NGC\,4253, NGC\,4748, IC\,4218, MCG-06-30-15, NGC\,5548}
\tablehead{
\multicolumn{2}{c}{NGC\,4253} &
\multicolumn{2}{c}{NGC\,4748} &
\multicolumn{2}{c}{IC\,4218} &
\multicolumn{2}{c}{MCG-06-30-15} &
\multicolumn{2}{c}{NGC\,5548}\\
\colhead{HJD} &
\colhead{$f$(H$\beta$)} &
\colhead{HJD} &
\colhead{$f$(H$\beta$)} &
\colhead{HJD} &
\colhead{$f$(H$\beta$)} &
\colhead{HJD} &
\colhead{$f$(H$\beta$)} &
\colhead{HJD} &
\colhead{$f$(H$\beta$)}}

\startdata

4550.8014  & $1.985 \pm  0.017$ &  4550.8173 &  $1.987 \pm  0.017$ &  4550.9372   &  $0.263 \pm  0.014$ &   4550.9007   &  $0.773  \pm  0.031$  &  4550.8678  &  $3.341 \pm  0.052$ \\ 
4551.8338  & $1.948 \pm  0.017$ &  4551.8530 &  $1.978 \pm  0.017$ &  4551.9372   &  $0.234 \pm  0.013$ &   4551.8929   &  $0.771  \pm  0.031$  &  4551.8664  &  $3.386 \pm  0.052$ \\ 
4553.8298  & $2.031 \pm  0.017$ &  4553.8454 &  $1.983 \pm  0.017$ &  4556.8372   &  $0.267 \pm  0.014$ &   4553.8838   &  $0.772  \pm  0.031$  &  4553.8626  &  $3.372 \pm  0.052$ \\ 
4555.9872  & $1.858 \pm  0.031$ &  4556.8087 &  $1.887 \pm  0.017$ &  4558.8351   &  $0.271 \pm  0.015$ &   4556.8813   &  $0.780  \pm  0.032$  &  4556.8589  &  $3.308 \pm  0.051$ \\ 
4556.7926  & $1.997 \pm  0.017$ &  4557.8618 &  $1.913 \pm  0.017$ &  4560.8290   &  $0.260 \pm  0.014$ &   4557.9053   &  $0.866  \pm  0.035$  &  4557.8755  &  $3.344 \pm  0.052$ \\ 
4557.7073  & $2.097 \pm  0.018$ &  4558.8076 &  $1.953 \pm  0.017$ &  4561.8328   &  $0.232 \pm  0.013$ &   4558.8788   &  $0.830  \pm  0.034$  &  4558.8595  &  $3.181 \pm  0.049$ \\ 
4558.6802  & $2.094 \pm  0.018$ &  4560.8578 &  $1.965 \pm  0.017$ &  4562.8158   &  $0.237 \pm  0.013$ &   4560.8720   &  $0.888  \pm  0.036$  &  4559.9903  &  $3.282 \pm  0.051$ \\ 
4559.9827  & $2.138 \pm  0.018$ &  4561.8522 &  $1.930 \pm  0.017$ &  4566.8828   &  $0.211 \pm  0.011$ &   4561.8670   &  $0.683  \pm  0.028$  &  4560.8967  &  $3.277 \pm  0.051$ \\ 
4560.7659  & $2.109 \pm  0.018$ &  4562.8377 &  $1.894 \pm  0.017$ &  4567.8865   &  $0.204 \pm  0.011$ &   4562.8549   &  $0.696  \pm  0.028$  &  4562.9997  &  $3.398 \pm  0.053$ \\ 
4561.7998  & $2.040 \pm  0.018$ &  4564.8024 &  $2.046 \pm  0.018$ &  4568.8743   &  $0.228 \pm  0.012$ &   4566.8546   &  $0.799  \pm  0.032$  &  4564.9405  &  $3.761 \pm  0.557$ \\ 
4562.7806  & $2.113 \pm  0.018$ &  4566.8071 &  $2.012 \pm  0.018$ &  4569.9003   &  $0.239 \pm  0.013$ &   4567.8529   &  $0.808  \pm  0.033$  &  4566.9017  &  $3.509 \pm  0.054$ \\ 
4564.7862  & $2.124 \pm  0.018$ &  4567.8104 &  $2.010 \pm  0.018$ &  4570.8983   &  $0.219 \pm  0.012$ &   4568.8453   &  $0.836  \pm  0.034$  &  4567.8218  &  $3.495 \pm  0.054$ \\ 
4566.7916  & $2.070 \pm  0.018$ &  4568.8070 &  $2.058 \pm  0.018$ &  4573.7922   &  $0.251 \pm  0.014$ &   4569.8506   &  $0.870  \pm  0.035$  &  4568.8201  &  $3.594 \pm  0.056$ \\ 
4567.7949  & $2.081 \pm  0.018$ &  4569.7977 &  $2.083 \pm  0.018$ &  4581.8880   &  $0.188 \pm  0.010$ &   4570.8414   &  $0.822  \pm  0.033$  &  4569.8830  &  $3.628 \pm  0.056$ \\ 
4568.7896  & $2.082 \pm  0.018$ &  4570.8224 &  $2.020 \pm  0.018$ &  4582.8930   &  $0.234 \pm  0.013$ &   4572.8345   &  $0.852  \pm  0.034$  &  4570.8760  &  $3.605 \pm  0.056$ \\ 
4569.8729  & $2.050 \pm  0.018$ &  4572.8723 &  $2.100 \pm  0.018$ &  4583.8888   &  $0.209 \pm  0.011$ &   4573.8404   &  $0.872  \pm  0.035$  &  4572.9341  &  $3.539 \pm  0.055$ \\ 
4570.8633  & $2.054 \pm  0.018$ &  4573.7696 &  $2.122 \pm  0.019$ &  4584.8577   &  $0.183 \pm  0.010$ &   4575.8398   &  $0.729  \pm  0.029$  &  4573.8140  &  $3.548 \pm  0.055$ \\ 
4572.9156  & $2.019 \pm  0.018$ &  4581.8356 &  $2.187 \pm  0.019$ &  4585.7753   &  $0.163 \pm  0.009$ &   4581.8155   &  $0.775  \pm  0.031$  &  4575.9500  &  $3.386 \pm  0.081$ \\ 
4573.8639  & $2.058 \pm  0.018$ &  4582.7554 &  $2.150 \pm  0.019$ &  4587.8627   &  $0.189 \pm  0.010$ &   4582.8097   &  $0.832  \pm  0.034$  &  4581.9237  &  $3.479 \pm  0.054$ \\ 
4575.8961  & $2.049 \pm  0.018$ &  4583.7544 &  $2.160 \pm  0.019$ &  4588.8565   &  $0.200 \pm  0.011$ &   4583.8102   &  $0.834  \pm  0.034$  &  4582.9322  &  $3.567 \pm  0.055$ \\ 
4581.9089  & $2.000 \pm  0.017$ &  4584.7718 &  $2.146 \pm  0.019$ &  4589.8671   &  $0.205 \pm  0.011$ &   4587.8043   &  $0.789  \pm  0.032$  &  4583.9276  &  $3.511 \pm  0.054$ \\ 
4582.9155  & $2.036 \pm  0.017$ &  4585.7473 &  $2.115 \pm  0.019$ &  4590.8627   &  $0.194 \pm  0.010$ &   4589.8018   &  $0.761  \pm  0.031$  &  4584.8941  &  $3.649 \pm  0.057$ \\ 
4583.9107  & $2.008 \pm  0.017$ &  4587.7612 &  $2.173 \pm  0.019$ &  4591.8663   &  $0.186 \pm  0.010$ &   4590.8007   &  $0.837  \pm  0.034$  &  4585.9384  &  $3.746 \pm  0.058$ \\ 
4584.8792  & $2.028 \pm  0.017$ &  4588.7502 &  $2.238 \pm  0.020$ &  4592.8623   &  $0.195 \pm  0.011$ &   4591.7959   &  $0.786  \pm  0.032$  &  4587.9254  &  $3.903 \pm  0.060$ \\ 
4585.8815  & $2.032 \pm  0.017$ &  4589.7575 &  $2.271 \pm  0.020$ &  4593.8630   &  $0.165 \pm  0.009$ &   4592.7947   &  $0.772  \pm  0.031$  &  4588.9167  &  $3.882 \pm  0.060$ \\ 
4587.8867  & $1.949 \pm  0.017$ &  4590.7558 &  $2.182 \pm  0.019$ &  4594.8339   &  $0.177 \pm  0.010$ &   4593.7952   &  $0.791  \pm  0.032$  &  4589.9298  &  $3.844 \pm  0.060$ \\ 
4588.8780  & $2.048 \pm  0.018$ &  4591.7470 &  $2.214 \pm  0.019$ &  4595.8284   &  $0.170 \pm  0.009$ &   4594.7825   &  $0.727  \pm  0.030$  &  4590.9214  &  $3.872 \pm  0.060$ \\ 
4589.8875  & $2.001 \pm  0.017$ &  4592.7507 &  $2.164 \pm  0.019$ &  4596.8309   &  $0.183 \pm  0.010$ &   4595.7851   &  $0.737  \pm  0.030$  &  4591.9278  &  $3.766 \pm  0.058$ \\ 
4590.8825  & $2.025 \pm  0.017$ &  4593.7517 &  $2.160 \pm  0.019$ &  4597.8582   &  $0.167 \pm  0.009$ &   4596.7772   &  $0.815  \pm  0.033$  &  4592.9217  &  $3.764 \pm  0.058$ \\ 
4591.8896  & $1.980 \pm  0.017$ &  4594.7569 &  $2.231 \pm  0.020$ &  4601.8491   &  $0.224 \pm  0.013$ &   4597.7787   &  $0.832  \pm  0.034$  &  4593.9218  &  $3.807 \pm  0.059$ \\ 
4592.8844  & $1.977 \pm  0.017$ &  4595.7658 &  $2.253 \pm  0.020$ &  4604.8890   &  $0.173 \pm  0.009$ &   4598.7755   &  $0.735  \pm  0.030$  &  4594.9230  &  $3.599 \pm  0.056$ \\ 
4593.8831  & $1.928 \pm  0.017$ &  4596.7525 &  $2.281 \pm  0.020$ &  4605.8573   &  $0.165 \pm  0.090$ &   4600.7774   &  $0.834  \pm  0.034$  &  4595.9071  &  $3.710 \pm  0.057$ \\ 
4594.8536  & $1.890 \pm  0.016$ &  4597.7547 &  $2.237 \pm  0.020$ &  4607.8548   &  $0.267 \pm  0.015$ &   4601.7573   &  $0.838  \pm  0.034$  &  4596.9126  &  $3.626 \pm  0.056$ \\ 
4595.8405  & $1.975 \pm  0.017$ &  4598.7508 &  $2.264 \pm  0.020$ &  4608.8580   &  $0.193 \pm  0.016$ &   4602.7605   &  $0.890  \pm  0.036$  &  4597.9181  &  $3.535 \pm  0.055$ \\ 
4596.8745  & $1.887 \pm  0.016$ &  4600.7957 &  $2.258 \pm  0.020$ &  4611.8449   &  $0.231 \pm  0.013$ &   4603.7690   &  $0.689  \pm  0.028$  &  4598.8420  &  $3.475 \pm  0.090$ \\ 
4597.8779  & $1.932 \pm  0.017$ &  4601.7432 &  $2.134 \pm  0.019$ &  4612.8694   &  $0.230 \pm  0.012$ &   4604.7505   &  $0.792  \pm  0.032$  &  4600.8564  &  $3.301 \pm  0.051$ \\ 
4598.8237  & $1.602 \pm  0.148$ &  4602.7264 &  $2.103 \pm  0.118$ &  4613.8330   &  $0.294 \pm  0.116$ &   4605.7525   &  $0.824  \pm  0.133$  &  4601.8688  &  $3.234 \pm  0.050$ \\
4600.7503  & $1.906 \pm  0.016$ &  4603.8304 &  $2.305 \pm  0.020$ &  4615.8112   &  $0.279 \pm  0.015$ &   4607.7576   &  $1.097  \pm  0.101$  &  4602.9239  &  $3.274 \pm  0.051$ \\ 
4601.7961  & $1.931 \pm  0.017$ &  4604.8659 &  $2.172 \pm  0.019$ &  4616.8177   &  $0.258 \pm  0.014$ &   4608.7375   &  $0.763  \pm  0.031$  &  4603.9194  &  $3.124 \pm  0.053$ \\ 
4602.7378  & $1.954 \pm  0.017$ &  4605.8354 &  $2.209 \pm  0.019$ &  4617.8407   &  $0.259 \pm  0.014$ &   4613.7611   &  $0.780  \pm  0.061$  &  4604.9353  &  $3.225 \pm  0.050$ \\ 
4603.8652  & $1.972 \pm  0.017$ &  4607.7997 &  $2.073 \pm  0.073$ &              &                     &   4614.7309   &  $0.826  \pm  0.033$  &  4605.9149  &  $3.064 \pm  0.047$ \\ 
4605.8768  & $1.934 \pm  0.017$ &  4608.7163 &  $2.085 \pm  0.018$ &     	  &  	                &   4615.7294   &  $0.827  \pm  0.033$  &  4607.8999  &  $3.131 \pm  0.048$ \\
4607.8742  & $1.862 \pm  0.016$ &  4614.7131 &  $2.097 \pm  0.018$ &              &  	                &      	        &                       &  4608.9063  &  $3.123 \pm  0.050$ \\
4608.8813  & $1.701 \pm  0.015$ &  4615.7067 &  $2.118 \pm  0.019$ &              &  	                &   	        &                       &  4611.9246  &  $2.968 \pm  0.796$ \\
4613.8473  & $2.053 \pm  0.018$ &  4618.7451 &  $2.075 \pm  0.018$ &              &  	                &  	        &                       &  4612.9069  &  $2.770 \pm  0.043$ \\
4614.8695  & $1.988 \pm  0.017$ &            &                     &              &   	                &  	        &                       &  4613.9068  &  $2.492 \pm  0.553$ \\
4615.8665  & $1.966 \pm  0.017$ &            &                     &              &   	                &  	        &                       &  4614.9259  &  $2.771 \pm  0.043$ \\
4616.8365  & $1.935 \pm  0.017$ &            &                     &              &   	                &  	        &                       &  4615.9124  &  $3.043 \pm  0.047$ \\
4617.8594  & $1.921 \pm  0.016$ &            &                     &              &   	                &  	        &                       &  4616.9108  &  $2.949 \pm  0.046$ \\
4618.8351  & $2.157 \pm  0.143$ &            &                     &              &   	                &  	        &                       &  4617.9020  &  $2.990 \pm  0.046$ \\
	   &  	                &	     &	                   &	          &                     &  	        &      	                &  4618.8734  &  $2.793 \pm  0.091$ \\

\enddata	    	     
\tablecomments{HJD = Heliocentric Julian Day$-$2450000; H$\beta$
               emission-line fluxes are in units of $10^{-13}$ erg
               s$^{-1}$ cm$^{-2}$.}
\label{table:lc2}
\end{deluxetable}


\begin{deluxetable}{cccccc}
\tablecolumns{6}
\tablewidth{0pt}
\tablecaption{H$\beta$ Light Curves --- Mrk\,290, IC\,1198, NGC\,6814}
\tablehead{
\multicolumn{2}{c}{Mrk\,290} &
\multicolumn{2}{c}{IC\,1198} &
\multicolumn{2}{c}{NGC\,6814} \\
\colhead{HJD} &
\colhead{$f$(H$\beta$)} &
\colhead{HJD} &
\colhead{$f$(H$\beta$)} &
\colhead{HJD} &
\colhead{$f$(H$\beta$)}}

\startdata

4550.9734   &  $3.438 \pm   0.061$ &  4551.0023   &  $1.186 \pm  0.019$ &  4551.0180  &  $2.987 \pm  0.030$  \\
4551.9662   &  $3.383 \pm   0.060$ &  4552.0039   &  $1.145 \pm  0.018$ &  4552.0264  &  $2.693 \pm  0.027$  \\
4555.9998   &  $3.401 \pm   0.060$ &  4556.0233   &  $1.163 \pm  0.018$ &  4556.0339  &  $2.562 \pm  0.035$  \\
4556.9015   &  $3.442 \pm   0.061$ &  4556.9309   &  $1.190 \pm  0.019$ &  4557.0200  &  $2.956 \pm  0.030$  \\
4558.8995   &  $3.443 \pm   0.061$ &  4558.9231   &  $1.190 \pm  0.019$ &  4560.0402  &  $2.924 \pm  0.037$  \\
4559.9990   &  $3.399 \pm   0.060$ &  4560.0230   &  $1.172 \pm  0.018$ &  4561.0266  &  $3.042 \pm  0.031$  \\
4560.9107   &  $3.348 \pm   0.030$ &  4560.9417   &  $1.175 \pm  0.019$ &  4564.0146  &  $3.096 \pm  0.032$  \\
4563.0117   &  $3.253 \pm   0.058$ &  4566.9602   &  $1.180 \pm  0.019$ &  4567.0021  &  $3.266 \pm  0.033$  \\
4563.9912   &  $3.396 \pm   0.060$ &  4567.9748   &  $1.136 \pm  0.018$ &  4568.0082  &  $3.138 \pm  0.032$  \\
4566.9350   &  $3.346 \pm   0.060$ &  4568.9233   &  $1.158 \pm  0.018$ &  4569.0070  &  $3.160 \pm  0.032$  \\
4567.9476   &  $3.248 \pm   0.058$ &  4570.9537   &  $1.159 \pm  0.018$ &  4570.0003  &  $3.139 \pm  0.032$  \\
4568.8957   &  $3.337 \pm   0.059$ &  4572.9665   &  $1.151 \pm  0.018$ &  4570.9899  &  $3.127 \pm  0.032$  \\
4569.9151   &  $3.284 \pm   0.058$ &  4573.9507   &  $1.145 \pm  0.018$ &  4572.9883  &  $3.173 \pm  0.032$  \\
4570.9189   &  $3.307 \pm   0.059$ &  4575.9921   &  $1.071 \pm  0.017$ &  4573.9952  &  $3.027 \pm  0.031$  \\
4572.9434   &  $3.269 \pm   0.058$ &  4581.9628   &  $1.148 \pm  0.018$ &  4576.0113  &  $3.109 \pm  0.032$  \\
4573.9102   &  $3.273 \pm   0.058$ &  4582.9746   &  $1.151 \pm  0.018$ &  4581.9978  &  $2.713 \pm  0.028$  \\
4575.9693   &  $3.189 \pm   0.057$ &  4583.9707   &  $1.156 \pm  0.018$ &  4583.0102  &  $2.695 \pm  0.027$  \\
4581.9349   &  $3.131 \pm   0.056$ &  4584.9733   &  $1.147 \pm  0.018$ &  4584.0028  &  $2.555 \pm  0.026$  \\
4582.9459   &  $3.204 \pm   0.057$ &  4585.9815   &  $1.108 \pm  0.017$ &  4585.0027  &  $2.544 \pm  0.026$  \\
4583.9414   &  $3.232 \pm   0.057$ &  4587.9626   &  $1.116 \pm  0.018$ &  4586.0112  &  $2.438 \pm  0.025$  \\
4584.9398   &  $3.111 \pm   0.055$ &  4588.9532   &  $1.165 \pm  0.018$ &  4588.0005  &  $2.571 \pm  0.026$  \\
4585.9516   &  $3.168 \pm   0.056$ &  4589.9672   &  $1.142 \pm  0.018$ &  4588.9927  &  $2.646 \pm  0.027$  \\
4587.9360   &  $3.101 \pm   0.055$ &  4590.9588   &  $1.165 \pm  0.018$ &  4590.0043  &  $2.627 \pm  0.027$  \\
4588.9259   &  $3.174 \pm   0.056$ &  4591.9632   &  $1.123 \pm  0.018$ &  4590.9920  &  $2.745 \pm  0.028$  \\
4589.9395   &  $3.157 \pm   0.056$ &  4592.9575   &  $1.149 \pm  0.018$ &  4591.9933  &  $2.602 \pm  0.026$  \\
4590.9318   &  $3.175 \pm   0.056$ &  4593.9851   &  $1.096 \pm  0.017$ &  4592.9872  &  $2.687 \pm  0.027$  \\
4591.9369   &  $3.153 \pm   0.056$ &  4594.9590   &  $1.141 \pm  0.018$ &  4593.9387  &  $2.627 \pm  0.027$  \\
4592.9306   &  $3.235 \pm   0.058$ &  4595.9419   &  $1.150 \pm  0.018$ &  4594.9950  &  $2.816 \pm  0.029$  \\
4593.9584   &  $3.196 \pm   0.057$ &  4596.9490   &  $1.184 \pm  0.019$ &  4595.9770  &  $2.815 \pm  0.029$  \\
4594.9322   &  $3.186 \pm   0.057$ &  4597.9552   &  $1.151 \pm  0.018$ &  4596.9783  &  $2.867 \pm  0.029$  \\
4595.9160   &  $3.265 \pm   0.058$ &  4598.9350   &  $0.989 \pm  0.146$ &  4597.9865  &  $2.941 \pm  0.030$  \\
4596.9218   &  $3.286 \pm   0.058$ &  4600.8938   &  $1.144 \pm  0.018$ &  4598.9686  &  $2.711 \pm  0.131$  \\
4597.9287   &  $3.168 \pm   0.056$ &  4601.9382   &  $1.102 \pm  0.017$ &  4600.9250  &  $2.999 \pm  0.031$  \\
4598.8656   &  $3.084 \pm   0.055$ &  4602.9611   &  $1.115 \pm  0.018$ &  4601.9789  &  $2.911 \pm  0.030$  \\
4600.8656   &  $3.196 \pm   0.057$ &  4603.9591   &  $1.106 \pm  0.017$ &  4602.9915  &  $3.027 \pm  0.031$  \\
4601.9111   &  $3.161 \pm   0.056$ &  4604.9686   &  $1.075 \pm  0.017$ &  4603.9917  &  $3.018 \pm  0.031$  \\
4602.9344   &  $3.177 \pm   0.157$ &  4605.9503   &  $1.116 \pm  0.018$ &  4604.9922  &  $3.074 \pm  0.031$  \\
4603.9305   &  $3.237 \pm   0.058$ &  4608.9364   &  $1.091 \pm  0.017$ &  4605.9838  &  $2.970 \pm  0.030$  \\
4604.9449   &  $3.235 \pm   0.058$ &  4612.9481   &  $1.114 \pm  0.018$ &  4608.9747  &  $2.639 \pm  0.027$  \\
4605.9238   &  $3.235 \pm   0.058$ &  4613.9446   &  $0.991 \pm  0.016$ &  4612.9746  &  $2.469 \pm  0.025$  \\
4607.9123   &  $3.284 \pm   0.058$ &  4614.9681   &  $1.114 \pm  0.018$ &  4613.9797  &  $2.066 \pm  0.136$  \\
4612.9202   &  $3.239 \pm   0.058$ &  4615.9499   &  $1.153 \pm  0.018$ &  4614.9871  &  $2.522 \pm  0.026$  \\
4613.9228   &  $3.143 \pm   0.056$ &  4616.9462   &  $1.063 \pm  0.017$ &  4616.9777  &  $2.600 \pm  0.026$  \\
4614.9427   &  $3.241 \pm   0.058$ &  4617.9372   &  $1.159 \pm  0.018$ &  4617.9693  &  $2.332 \pm  0.024$  \\
4615.9226   &  $3.313 \pm   0.059$ &  4618.9607   &  $1.203 \pm  0.019$ &  4618.9885  &  $2.585 \pm  0.026$  \\
4616.9199   &  $3.154 \pm   0.056$ &  	          &  	                &	      &                      \\
4617.9115   &  $3.407 \pm   0.061$ &  	          &  	                &	      &        	             \\
4618.9014   &  $3.415 \pm   0.061$ &  	          &  	                &             &                      \\

\enddata													
\tablecomments{HJD = Heliocentric Julian Day$-$2450000; H$\beta$
               emission-line fluxes are in units of $10^{-13}$ erg
               s$^{-1}$ cm$^{-2}$.}
\label{table:lc3}
\end{deluxetable}    
	     
		     
\begin{deluxetable}{lccccccc}
\tablecolumns{8}
\tablewidth{0pt}
\tablecaption{Light-Curve Statistics}
\tablehead{
\colhead{Object} &
\colhead{Time Series} &
\colhead{$N$} &
\colhead{$\langle T \rangle$} &
\colhead{$T_{\rm median}$} &
\colhead{$\langle \sigma_{\rm f}/f \rangle$} &
\colhead{$F_{\rm var}$} &
\colhead{$R_{\rm max}$}\\
\colhead{(1)} &
\colhead{(2)} &
\colhead{(3)} &
\colhead{(4)} &
\colhead{(5)} &
\colhead{(6)} &
\colhead{(7)} &
\colhead{(8)} }
\startdata

Mrk\,142       & $B$       & 64 & $1.8 \pm 2.3$ & 1.02 & 0.0166 & 0.025 & $1.15 \pm 0.03$ \\
	       & $V$       & 62 & $1.7 \pm 2.0$ & 1.02 & 0.0119 & 0.024 & $1.12 \pm 0.02$ \\
	       & H$\beta$  & 51 & $1.4 \pm 1.0$ & 1.00 & 0.0113 & 0.086 & $1.97 \pm 0.03$ \\
SBS\,1116+583A & $B$       & 56 & $2.1 \pm 1.8$ & 1.02 & 0.0205 & 0.104 & $1.63 \pm 0.05$ \\
               & $V$       & 56 & $1.9 \pm 1.7$ & 1.01 & 0.0220 & 0.082 & $1.47 \pm 0.05$ \\
               & H$\beta$  & 50 & $1.4 \pm 0.9$ & 1.00 & 0.0279 & 0.102 & $1.48 \pm 0.06$ \\
Arp\,151       & $B$       & 66 & $1.5 \pm 1.6$ & 1.02 & 0.0173 & 0.161 & $1.80 \pm 0.04$ \\
               & $V$       & 62 & $1.6 \pm 1.6$ & 1.02 & 0.0185 & 0.113 & $1.54 \pm 0.04$ \\
               & H$\beta$  & 43 & $1.4 \pm 1.9$ & 1.02 & 0.0153 & 0.169 & $1.74 \pm 0.04$ \\
Mrk\,1310      & $B$       & 50 & $2.0 \pm 1.5$ & 1.16 & 0.0160 & 0.116 & $1.71 \pm 0.04$ \\
	       & $V$       & 58 & $1.8 \pm 1.4$ & 1.05 & 0.0183 & 0.073 & $1.39 \pm 0.04$ \\
	       & H$\beta$  & 47 & $1.5 \pm 1.1$ & 1.01 & 0.0186 & 0.108 & $1.62 \pm 0.04$ \\
Mrk\,202       & $B$       & 58 & $2.0 \pm 1.7$ & 1.01 & 0.0168 & 0.042 & $1.20 \pm 0.03$ \\
	       & $V$       & 58 & $1.8 \pm 1.7$ & 1.01 & 0.0143 & 0.027 & $1.18 \pm 0.04$ \\
 	       & H$\beta$  & 46 & $1.5 \pm 1.2$ & 1.01 & 0.0125 & 0.089 & $1.42 \pm 0.03$ \\
NGC\,4253      & $B$       & 51 & $1.9 \pm 2.3$ & 1.02 & 0.0066 & 0.032 & $1.16 \pm 0.01$ \\
	       & $V$       & 54 & $1.8 \pm 2.2$ & 1.01 & 0.0046 & 0.028 & $1.15 \pm 0.01$ \\
	       & H$\beta$  & 50 & $1.4 \pm 1.0$ & 1.01 & 0.0116 & 0.048 & $1.35 \pm 0.15$ \\
NGC\,4748      & $B$       & 48 & $2.4 \pm 3.1$ & 1.25 & 0.0151 & 0.053 & $1.22 \pm 0.05$ \\
	       & $V$       & 52 & $2.2 \pm 2.5$ & 1.03 & 0.0147 & 0.043 & $1.18 \pm 0.02$ \\
	       & H$\beta$  & 45 & $1.5 \pm 1.3$ & 1.00 & 0.0094 & 0.052 & $1.22 \pm 0.02$ \\
IC\,4218       & $B$       & 42 & $2.8 \pm 5.5$ & 1.03 & 0.0154 & 0.087 & $1.42 \pm 0.04$ \\
	       & $V$       & 65 & $2.1 \pm 1.9$ & 1.09 & 0.0203 & 0.079 & $1.52 \pm 0.06$ \\
	       & H$\beta$  & 40 & $1.7 \pm 1.5$ & 1.00 & 0.0551 & 0.159 & $1.90 \pm 0.14$ \\
MCG-06-30-15   & $B$       & 48 & $2.1 \pm 1.7$ & 1.08 & 0.0165 & 0.037 & $1.19 \pm 0.03$ \\
	       & $V$       & 55 & $1.9 \pm 1.6$ & 1.04 & 0.0192 & 0.032 & $1.21 \pm 0.03$ \\
	       & H$\beta$  & 42 & $1.6 \pm 1.2$ & 1.00 & 0.0442 & 0.067 & $1.61 \pm 0.26$ \\
NGC\,5548      & $B$       & 45 & $2.4 \pm 4.3$ & 1.07 & 0.0148 & 0.085 & $1.39 \pm 0.03$ \\
	       & $V$       & 57 & $1.9 \pm 1.8$ & 1.05 & 0.0125 & 0.094 & $1.40 \pm 0.02$ \\
	       & H$\beta$  & 51 & $1.4 \pm 0.9$ & 1.01 & 0.0279 & 0.082 & $1.57 \pm 0.35$ \\
Mrk\,290       & $B$       & 50 & $2.3 \pm 2.8$ & 1.01 & 0.0107 & 0.038 & $1.23 \pm 0.02$ \\
	       & $V$       & 50 & $2.1 \pm 2.5$ & 1.01 & 0.0107 & 0.024 & $1.12 \pm 0.02$ \\
	       & H$\beta$  & 48 & $1.5 \pm 1.1$ & 1.01 & 0.0178 & 0.025 & $1.12 \pm 0.03$ \\
IC\,1198       & $B$       & 55 & $2.0 \pm 2.0$ & 1.05 & 0.0185 & 0.039 & $1.21 \pm 0.03$ \\
	       & $V$       & 58 & $1.6 \pm 1.7$ & 1.02 & 0.0134 & 0.031 & $1.16 \pm 0.02$ \\
	       & H$\beta$  & 45 & $1.5 \pm 1.2$ & 1.01 & 0.0187 & 0.031 & $1.22 \pm 0.18$ \\
NGC\,6814      & $B$       & 43 & $1.7 \pm 1.3$ & 1.04 & 0.0137 & 0.178 & $1.83 \pm 0.03$ \\
	       & $V$       & 46 & $1.6 \pm 1.3$ & 1.02 & 0.0134 & 0.145 & $1.68 \pm 0.03$ \\
	       & H$\beta$  & 45 & $1.5 \pm 1.1$ & 1.01 & 0.0124 & 0.093 & $1.58 \pm 0.11$ \\
\enddata 

\tablecomments{Columns are presented as follows: (1) object; (2)
               feature; (3) number of observations; (4) average
               interval between observations in days; (5) median
               sampling rate in days; (6) mean fractional error; (7)
               excess variance as described in the text; and (8) the
               ratio of the maximum to the minimum flux.}
\label{table:variability}
\end{deluxetable}


\begin{deluxetable}{lcccc}
\tablecolumns{5}
\tablewidth{0pt}
\tablecaption{H$\beta$ Time Lag Measurements}
\tablehead{
\colhead{} &
\multicolumn{2}{c}{Observed} &
\multicolumn{2}{c}{Rest-frame} \\
\colhead{Object} &
\colhead{$\tau_{\rm cent}$} &
\colhead{$\tau_{\rm peak}$} &
\colhead{$\tau_{\rm cent}$} &
\colhead{$\tau_{\rm peak}$} \\
\colhead{} &
\colhead{(days)} &
\colhead{(days)} &
\colhead{(days)} &
\colhead{(days)}} 
\startdata

\multicolumn{5}{c}{vs. $B$ band} \\ \hline \\

Mrk\,142       & $2.87^{+0.76}_{-0.87}$ & $2.75^{+1.00}_{-0.75}$ & $2.74^{+0.73}_{-0.83}$ & $2.63^{+0.96}_{-0.72}$ \\
SBS\,1116+583A & $2.38^{+0.64}_{-0.51}$ & $2.25^{+1.00}_{-0.50}$ & $2.31^{+0.62}_{-0.49}$ & $2.19^{+0.97}_{-0.49}$ \\
Arp\,151       & $4.08^{+0.50}_{-0.69}$ & $3.50^{+0.75}_{-0.25}$ & $3.99^{+0.49}_{-0.68}$ & $3.43^{+0.73}_{-0.24}$ \\
Mrk\,1310      & $3.74^{+0.60}_{-0.62}$ & $3.75^{+0.50}_{-0.50}$ & $3.66^{+0.59}_{-0.61}$ & $3.68^{+0.49}_{-0.49}$ \\
Mrk\,202       & $3.12^{+1.77}_{-1.15}$ & $3.00^{+1.50}_{-1.25}$ & $3.05^{+1.73}_{-1.12}$ & $2.94^{+1.47}_{-1.22}$ \\
NGC\,4253      & $6.24^{+1.65}_{-1.24}$ & $6.00^{+2.50}_{-1.00}$ & $6.16^{+1.63}_{-1.22}$ & $5.92^{+2.47}_{-0.99}$ \\
NGC\,4748      & $5.63^{+1.64}_{-2.25}$ & $5.75^{+3.50}_{-2.00}$ & $5.55^{+1.62}_{-2.22}$ & $5.67^{+3.45}_{-1.97}$ \\
NGC\,5548      & $4.25^{+0.88}_{-1.33}$ & $4.25^{+1.25}_{-1.50}$ & $4.18^{+0.86}_{-1.30}$ & $4.18^{+1.23}_{-1.47}$ \\
NGC\,6814      & $6.67^{+0.88}_{-0.90}$ & $7.25^{+0.25}_{-0.75}$ & $6.64^{+0.87}_{-0.90}$ & $7.21^{+0.25}_{-0.75}$ \\

\\
\hline \\
\multicolumn{5}{c}{vs. $V$ band} \\ \hline \\

Mrk\,142       & $2.88^{+1.00}_{-1.01}$ & $3.25^{+0.75}_{-1.75}$ & $2.76^{+0.96}_{-0.96}$ & $3.11^{+0.72}_{-1.67}$ \\
SBS\,1116+583A & $2.24^{+0.65}_{-0.61}$ & $2.25^{+0.75}_{-0.50}$ & $2.18^{+0.63}_{-0.60}$ & $2.19^{+0.73}_{-0.49}$ \\
Arp\,151       & $3.52^{+0.82}_{-0.72}$ & $3.50^{+1.00}_{-0.75}$ & $3.45^{+0.80}_{-0.71}$ & $3.43^{+0.98}_{-0.73}$ \\
Mrk\,1310      & $3.67^{+0.46}_{-0.50}$ & $3.75^{+0.50}_{-0.50}$ & $3.60^{+0.45}_{-0.49}$ & $3.68^{+0.49}_{-0.49}$ \\
Mrk\,202       & $3.11^{+0.91}_{-1.12}$ & $2.75^{+1.75}_{-1.25}$ & $3.05^{+0.89}_{-1.10}$ & $2.69^{+1.71}_{-1.22}$ \\
NGC\,4253      & $6.87^{+1.22}_{-1.84}$ & $6.50^{+2.25}_{-2.00}$ & $6.78^{+1.20}_{-1.81}$ & $6.42^{+2.22}_{-1.97}$ \\
NGC\,4748      & $6.39^{+1.84}_{-1.46}$ & $7.75^{+1.75}_{-3.75}$ & $6.30^{+1.82}_{-1.44}$ & $7.64^{+1.72}_{-3.70}$ \\
NGC\,5548      & $4.24^{+0.91}_{-1.35}$ & $4.25^{+1.50}_{-1.25}$ & $4.17^{+0.90}_{-1.33}$ & $4.18^{+1.47}_{-1.23}$ \\
NGC\,6814      & $6.49^{+0.95}_{-0.96}$ & $7.00^{+0.50}_{-0.50}$ & $6.46^{+0.94}_{-0.96}$ & $6.96^{+0.50}_{-0.50}$ \\

\enddata
\label{table:tau}
\end{deluxetable}


\begin{deluxetable}{lcc}
\tablecolumns{3}
\tablewidth{0pt}
\tablecaption{H$\beta$ Narrow-Component Strength}
\tablehead{
\colhead{Object} &
\colhead{$f$(H$\beta$)/$f$([\ion{O}{3}] $\lambda 5007$)} &
\colhead{Ref.} }
\startdata

Mrk\,142        & 0.274 & 1 \\
SBS\,1116+583A  & 0.07  & 2 \\
Arp\,151        & 0.15  & 2 \\
Mrk\,1310       & 0.13  & 2 \\
Mrk\,202        & 0.30  & 2 \\
NGC\,4253       & 0.113 & 1 \\
NGC\,4748       & 0.15  & 2 \\
NGC\,5548       & 0.114 & 1 \\
NGC\,6814       & 0.03  & 2 \\

\enddata

\tablerefs{ 1. \citet{marziani03}, 2. This work.}
\label{table:ratio}
\end{deluxetable}


\begin{deluxetable}{lcc}
\tablecolumns{3}
\tablewidth{0pt}
\tablecaption{[\ion{O}{3}] $\lambda 5007$ Line Widths and Spectral Resolution}
\tablehead{
\colhead{Object} &
\colhead{FWHM ([\ion{O}{3}] $\lambda 5007$)} &
\colhead{$\Delta \lambda_{\rm res}$} \\
\colhead{} &
\colhead{(km s$^{-1}$)} &
\colhead{(\AA)} }
\startdata

Arp\,151   & 220 & 13.1 \\
Mrk\,1310  & 120 & 12.4 \\
NGC\,4253  & 180 & 14.6 \\
NGC\,5548  & 410 & 14.7 \\
Mrk\,290   & 380 & 11.6 \\
IC\,1198   & 280 & 12.0 \\
NGC\,6814  & 125 & 12.9 \\

\enddata

\tablecomments{ \mbox{Line} widths are from \citet{whittle92}.}
\label{table:o3width}
\end{deluxetable}


\begin{deluxetable}{lcccc}
\tablecolumns{5}
\tablewidth{0pt}
\tablecaption{Rest-frame Broad H$\beta$ Line-Width Measurements}
\tablehead{
\colhead{} &
\multicolumn{2}{c}{Mean Spectrum} &
\multicolumn{2}{c}{Rms Spectrum} \\
\colhead{Object} &
\colhead{$\sigma_{\rm line}$} &
\colhead{FWHM} &
\colhead{$\sigma_{\rm line}$} &
\colhead{FWHM}\\
\colhead{} &
\colhead{(km s$^{-1}$)} &
\colhead{(km s$^{-1}$)} &
\colhead{(km s$^{-1}$)} &
\colhead{(km s$^{-1}$)} }
\startdata

Mrk\,142        & $1116 \pm 22$ & $1462  \pm 2$    & $859  \pm 102$ & $1368  \pm 379$  \\
SBS\,1116+583A  & $1552 \pm 36$ & $3668  \pm 186$  & $1528 \pm 184$ & $3604  \pm 1123$ \\
Arp\,151        & $2006 \pm 24$ & $3098  \pm 69$   & $1252 \pm 46$  & $2357  \pm 142$  \\
Mrk\,1310       & $1209 \pm 42$ & $2409  \pm 24$   & $755  \pm 138$ & $1602  \pm 250$  \\
Mrk\,202        & $867  \pm 40$ & $1471  \pm 18$   & $659  \pm 65$  & $1354  \pm 250$  \\
NGC\,4253       & $1088 \pm 37$ & $1609  \pm 39$   & $516  \pm 218$ & $834   \pm 1260$ \\
NGC\,4748       & $1009 \pm 27$ & $1947  \pm 66$   & $657  \pm 91$  & $1212  \pm 173$  \\
NGC\,5548       & $4266 \pm 65$ & $12771 \pm 71$   & $4270 \pm 292$ & $11177 \pm 2266$ \\
NGC\,6814       & $1918 \pm 36$ & $3323  \pm 7$    & $1610 \pm 108$ & $3277  \pm 297$  \\ 

\enddata
\label{table:hbwidth}
\end{deluxetable}


\begin{deluxetable}{lcc}
\tablecolumns{3}
\tablewidth{0pt}
\tablecaption{Virial Products and Derived Black Hole Masses}
\tablehead{
\colhead{Object} &
\colhead{$c \tau_{\rm cent} \sigma^2_{\rm line} / G$} &
\colhead{\mbh\tablenotemark{a}}\\
\colhead{} &
\colhead{($10^6$ M$_{\odot}$)} &
\colhead{($10^6$ M$_{\odot}$)} }
\startdata

Mrk\,142        & $0.40^{+0.14}_{-0.15}$ & $2.17^{+0.77}_{-0.83}$ \\ 
SBS\,1116+583A  & $1.05^{+0.38}_{-0.34}$ & $5.80^{+2.09}_{-1.86}$ \\ 
Arp\,151        & $1.22^{+0.17}_{-0.23}$ & $6.72^{+0.96}_{-1.24}$ \\ 
Mrk\,1310       & $0.41^{+0.16}_{-0.16}$ & $2.24^{+0.90}_{-0.90}$ \\ 
Mrk\,202        & $0.26^{+0.16}_{-0.11}$ & $1.42^{+0.85}_{-0.59}$ \\ 
NGC\,4253       & $0.32^{+0.28}_{-0.25}$ & $1.76^{+1.56}_{-1.40}$ \\ 
NGC\,4748       & $0.47^{+0.19}_{-0.23}$ & $2.57^{+1.03}_{-1.25}$ \\ 
NGC\,5548       & $14.9^{+3.7 }_{-5.1 }$ & $82^{+20}_{-28}$       \\ 
NGC\,6814       & $3.36^{+0.63}_{-0.64}$ & $18.5^{+3.5 }_{-3.5 }$ \\ 

\enddata

\tablenotetext{a}{Assuming $f=5.5$.}
\label{table:mbh}
\end{deluxetable}

\end{document}